\documentclass[aps,nofootinbib,superscriptaddress, showpacs,preprintnumbers,nofootinbibt,twocolumn]{revtex4-2}
\usepackage{epsfig}
\usepackage{multirow}
\usepackage{eurosym}
\usepackage{dcolumn}
\usepackage{bm}
\usepackage{enumerate}
\usepackage{float}
\usepackage{epstopdf}
\usepackage{amsmath}
\usepackage{bm}
\usepackage{amsfonts}
\usepackage{amssymb}
\usepackage{graphicx}
\usepackage{alphalph,mathtools}
\usepackage{etoolbox}
\usepackage{color}
\usepackage{booktabs}
\usepackage{lineno}
\usepackage{footnote}
\usepackage{makecell,tabularx}
\usepackage{hyperref}
\hypersetup{colorlinks,citecolor=blue}
\hypersetup{colorlinks=true,linkcolor=magenta,filecolor=magenta,
    urlcolor=blue}
    
\def\be{\begin{equation}}
    \def\ee{\end{equation}}
\def\bea{\begin{eqnarray}}
    \def\eea{\end{eqnarray}}

\begin{document} 


\title{Strong Gravitational Lensing by Sgr A* and M87* Black Holes embedded in Dark Matter Halo exhibiting string cloud and quintessential field}

\author{Niyaz Uddin Molla}
\email{niyazuddin182@gmail.com} 
\affiliation{Department of
Mathematics, Indian Institute of Engineering Science and
Technology, Shibpur, Howrah-711 103, India,}
\author{Himanshu Chaudhary}
\email{himanshuch1729@gmail.com} 
\affiliation{Pacif Institute of Cosmology and Selfology (PICS), Sagara, Sambalpur 768224, Odisha, India}
\affiliation{Department of Applied Mathematics, Delhi Technological University, Delhi-110042, India,}
\affiliation{Department of Mathematics, Shyamlal College, University of Delhi, Delhi-110032, India,}
\author{Dhruv Arora}
\email{arora09dhruv@gmail.com}
\affiliation{Pacif Institute of Cosmology and Selfology (PICS), Sagara, Sambalpur 768224, Odisha, India}
\author{Farruh Atamurotov}
\email{atamurotov@yahoo.com} 
\affiliation{Inha University in  
Tashkent, Ziyolilar 9, Tashkent 100170, Uzbekistan}
\affiliation{Central Asian University, Milliy Bog' Street 264, Tashkent 111221, Uzbekistan}
\affiliation{National University of Uzbekistan, Tashkent 100174, Uzbekistan}
\author{Ujjal Debnath}
\email{ujjaldebnath@gmail.com} \affiliation{Department of
Mathematics, Indian Institute of Engineering Science and
Technology, Shibpur, Howrah-711 103, India,}
\author{G.Mustafa}
\email{gmustafa3828@gmail.com}
\affiliation{Department of Physics,
Zhejiang Normal University, Jinhua 321004, People’s Republic of China}
\affiliation{New Uzbekistan University, Mustaqillik ave. 54, 100007 Tashkent, Uzbekistan}


\begin{abstract}
We investigate the strong gravitational lensing phenomena caused by a black hole with a dark matter halo. In this study, we examine strong gravitational lensing with two significant dark matter models: the universal rotation curve model and the cold dark matter model. To do this, we first numerically estimate the strong lensing coefficients and strong deflection angles for both the universal rotation curve and cold dark matter models. It is observed that the deflection angle, denoted as $\alpha_D$, increases with the parameter $\alpha$ while holding the value of $\gamma \cdot 2M$ constant. Additionally, it increases with the parameter $\gamma \cdot 2M$ while keeping the value of $\alpha$ constant. The strong deflection angle, $\alpha_D$, for the black hole with a dark matter halo, with parameters $\alpha=0$ and $\gamma \cdot 2M$, greatly enhances the gravitational bending effect and surpasses the corresponding case of the standard Schwarzschild black hole ($A=0, B=0, \alpha=0, \gamma \cdot 2M=0$). Furthermore, we investigate the astrophysical consequences through strong gravitational lensing observations, using examples of various supermassive black holes such as $M87^{*}$ and $SgrA^{*}$ located at the centers of several galaxies. It is observed that black holes with dark matter halos can be quantitatively distinguished and characterized from the standard Schwarzschild black hole ($A=0, B=0, \alpha=0, \gamma \cdot 2M=0$). The findings in our analysis suggest that observational tests for black holes influenced by dark matter halos are indeed feasible and viable.
\end{abstract}
\maketitle
\section{Introduction}\label{sec1}
The most fascinating prediction of general relativity is black holes, which became a theoretical possibility when Karl Schwarzschild first provided a solution to Einstein's field equations. This solution described the gravitational field outside a spherically symmetric, non-rotating, and static object in space \cite{schwarzschild2003gravitational}. In 1939, Oppenheimer, Snyder, and Datt \cite{oppenheimer1939continued} studied spherically symmetric and homogeneous dust collapse, concluding that it inevitably leads to the formation of Schwarzschild black holes. This provided the first possible astrophysical origin for black holes. Since then, Einstein's general theory of relativity has gained much popularity in recent years, passing multiple observational tests and establishing itself as the current best theory of gravity. Notably, the announcement of gravitational waves \cite{scientific2018erratum,scientific2017gw170104}, the first black hole shadow image of M87* and Sag A* by the EHT collaboration \cite{akiyama2019event,akiyama2022first,wielgus2022millimeter,akiyama2022first}, and the motion of S-stars around the galactic center \cite{do2019relativistic,abuter2018detection,abuter2020detection} have sparked much interest in the scientific community to study further the causal structure of spacetime and the dynamical environment around compact objects. One such effect of general relativity is the deflection of light by massive bodies. This leads to a phenomenon called gravitational lensing, first observed by the astronomer Walsh et al. in 1979 \cite{walsh19790957+}. It is currently an active research field with dedicated projects such as the Kilo--Degree survey \cite{li2023kids}, Gaia Gravitational Lenses group \cite{jablonska2022possible}, and the Strong-Lensing Insights into Dark Energy Survey \cite{taak2023strong}, among others. Our analysis in this paper focuses on strong gravitational lensing, which occurs when the source is almost exactly behind a very massive object, i.e., the lens. This phenomenon produces multiple images, arcs, and even Einstein rings. With improvements in observational astronomy, strong lensing has become of prime importance in testing various theories of gravity \cite{virbhadra2000schwarzschild,bozza2001strong,bozza2002gravitational,bozza2003quasiequatorial,chen2018new,zhang2017strong,zhang2018strong}.\\\\
Consequently, the theory of gravitational lensing can be classified into two distinct scenarios. In the first scenario, when a photon's radial distance greatly exceeds the gravitational lens's radius, the deflection of the light ray is very small. This results in two slightly distorted images on both sides of the lens. In the second scenario, referred to as strong gravitational lensing, the photon follows multiple orbits around the lens before reaching the observer. In this case, an infinite series of highly magnified images appears on both sides of the lens. The study of gravitational lensing in the vicinity of relativistic objects, such as a star orbiting in the Kerr spacetime \cite{cunningham1972optical} or an accretion disk surrounding a Schwarzschild black hole, was initially conducted in the 1970s \cite{Luminet:1979nyg}. The concept of gravitational lensing was first introduced in the weak field limit \cite{Refsdal:1964yk, Refsdal:1964nw, schneider1992gravitational, petters2001singularity}. It was later widely explored in the strong field limit \cite{darwin1959gravity, Claudel:2000yi, Bozza:2001xd, Bozza:2005tg}. Darwin conducted pioneering research on the paths of photons as they traverse the vicinity of a black hole, revealing the substantial bending of light rays. This work led to the development of a precise lens equation, as discussed later by Frittelli et al. in \cite{Frittelli:1999yf} and by Virbhadra $\&$ Ellis in 2000. Subsequently, in further studies, Virbhadra et al. \cite{Virbhadra:1998dy} and Virbhadra $\&$ Ellis \cite{Virbhadra:1999nm} embarked on numerical investigations, focusing on the phenomenon of lensing caused by static, spherically symmetric naked singularities. These investigations involved a meticulous examination of various lensing observables. Inspired by their work, Perlick \cite{Perlick:2003vg} delved into the study of lensing within spherically symmetric and static spacetime, employing the light-like geodesic equation without employing any approximations. The research has unveiled the profound significance of gravitational lensing, particularly in the context of black holes, as an invaluable tool in astrophysics. It allows for the exploration of the strong gravitational effects in the vicinity of these massive celestial objects and provides valuable insights into distant and faint stars. Gravitational lensing by various black holes has garnered considerable attention over recent decades due to its critical importance. Researchers have employed functional and analytical techniques, as introduced by Bozza et al. in their 2001 work \cite{Bozza:2001xd}, which are founded on the strong deflection of light rays. These techniques have revealed that the deflection angle exhibits a logarithmic divergence as light rays approach the photon sphere of a Schwarzschild black hole. Bozza's approach has not been limited to Schwarzschild black holes but has also been extended to other scenarios. For example, it has been applied to the study of Reissner–Nordström black holes \cite{Eiroa:2002mk} and implemented in metric systems that are both static and spherically symmetric \cite{Bozza:2002zj}. The extensive research on gravitational lensing is motivated by the understanding that the paths of light near black holes provide valuable insights into the fundamental features and characteristics of the underlying spacetime geometry. Numerous studies have been dedicated to examining gravitational lensing caused by a variety of sources, including the distribution of structures. \cite{Mellier:1998pk,Bartelmann:1999yn,Heymans:2013fya}, dark energy \cite{Biesiada:2006zf,DES:2020ahh,Huber:2021nhx}, dark matter (DM) \cite{Kaiser:1992ps,clowe2006direct,Atamurotov:2021hoq}, quasars \cite{SDSS:2000jpb,Peng:2006ew,Oguri:2010ns,yue2022revisiting} , gravitational waves \cite{Seljak:2003pn,Diego:2021fyd,finke2021probing}, and various other compact celestial objects \cite{Liao:2015uzb,Nascimento:2020ime,Junior:2021svb,Molla:2022mjl,Molla:2022izk}.  Recently, the Event Horizon Telescope collaboration successfully achieved an angular resolution that enables the observation of the image of the supermassive black hole situated at the center of the M87 galaxy. This development marks the onset of a new era in the exploration of gravitational lensing within the context of strong gravitational fields. \cite{EventHorizonTelescope:2019dse,EventHorizonTelescope:2019uob,EventHorizonTelescope:2019jan,EventHorizonTelescope:2019ths,EventHorizonTelescope:2019pgp,EventHorizonTelescope:2019ggy}.
Additionally, the anticipation that relativistic images can serve as a means to test gravity in the strong deflection limit has prompted the application of this technique to a wide range of black hole metrics. This includes both the context of general relativity and modified theories. Such an approach has led to diverse investigations involving various black hole metrics. This is evident from the contributions of researchers, such as Fernando and Roberts.\cite{fernando2002gravitational}, Bozza \cite{Bozza:2003cp}, Majumdar \& Mukherjee \cite{majumdar2005gravitational}, Eiroa \& Romero \cite{Eiroa:2008ks}, Bozza \cite{bozza2014gravitational}, Sahu et al. \cite{Sahu:2015dea}, and Islam \& Ghosh \cite{Islam:2021dyk}.\\\\
Furthermore, there has been considerable interest in studying gravitational lensing in the context of naked singularities or spacetimes without a horizon \cite{Gyulchev:2007cy}. The quantitative aspects of gravitational lensing have proven to be a valuable means of distinguishing between horizonless compact objects and black holes \cite{Gyulchev:2007cy,Gyulchev:2008ff}. In 2019, Shaikh et al \cite{Shaikh:2019itn}. conducted an analytical investigation into the phenomenon of strong gravitational lensing caused by such horizon-less compact objects. Their study aimed to derive precise expressions for lensing observables pertaining to images formed from within the photon sphere and then compared these results with images formed from outside the photon sphere. The astrophysical implications concerning various observables like image position, separation, magnification, the occurrence of Einstein rings, and time delays in the formation of relativistic images have been investigated in connection with several supermassive black holes \cite{Kumar:2019pjp,Kumar:2019ohr,Islam:2020xmy,Kumar:2020hgm,Kumar:2021cyl,Ghosh:2020spb,Guerrero2020RotatingBH,Kumar:2022fqo,2021PhRvD.103h4057B,Kumar:2020sag,Hsieh:2021rru,Molla:2023hou,Jha:2023qqz}. Recently, Influence of string cloud and quintessential field on optical properties of black hole is also considered in details in \cite{2022ChPhC..46l5107M,2022EPJC...82..831A,2023ChPhC..47b5102A}.\\\\ 
In this paper, we aim to expand upon the study of strong gravitational lensing by the supermassive black holes, Sgr A* and M87*, which are surrounded by DM halos exhibiting string clouds and quintessential fields. Our objective is to explore how this phenomenon can be distinguished from other astrophysical black holes. Additionally, we will compare the observable signatures of black holes in DM halos with string clouds and quintessential fields to those of Schwarzschild black holes. This comparison will take into consideration the supermassive black holes, Sgr A* and M87*, as the gravitational lenses.\\\\
The true nature of DM remains unclear to this day, making it a highly active research field in astrophysics, astronomy, and cosmology. Current indirect evidence for DM has been found through various means, including galactic rotation curves of spiral galaxies \cite{rubin1970rotation,corbelli2000extended}, gravitational lensing \cite{clowe2006direct}, large-scale structure formation \cite{davis1985evolution}, baryon acoustic oscillations (BAO), and cosmic microwave background radiation \cite{collaboration2014planck,bennett2011seven}. The rotation curves of spiral galaxies reveal that the velocity of stars in the outer regions of galaxies becomes constant \cite{persic1996universal}. This is in contrast to Newtonian dynamics, which predicts that the outermost stars must be rotating faster. The most logical explanation for this discrepancy is the presence of an unseen extra mass that causes such behavior. Observations from the cosmic microwave background show that 26.8 \% of our universe is composed of DM, while the remaining 68.3 \% is made up of dark energy \cite{collaboration2014planck}. The matter distribution in our galaxy is divided into different parts. The region from $10^{4}$ pc to $10^{5}$ pc is dominated by DM, often referred to as the outer halo \cite{krut2018galactic}. The most promising candidates for DM are particles beyond the standard model, such as Weakly Interacting Massive Particles (WIMP), axions, and dark photons \cite{bertone2005particle,boehm2004scalar,feng2009hidden,graham2015experimental}. Micro and primordial black holes \cite{bergstrom2009dark,abadaalaboratory} and sterile neutrinos are also among the possible candidates. Various modifications of Newton's theory of gravitation and the general theory of relativity have been proposed in the literature to explain the peculiar behavior of galactic rotation curves \cite{milgrom1983modification,mannheim1997galactic,roberts2004galactic,moffat1996galaxy,brownstein2006galaxy,mak2004can,harko2006galactic,boehmer2007galactic}.\\\\
The paper is structured as follows: In Section \ref{sec2}, we offer a concise description of black hole spacetime structure and DM halos. Section \ref{sec3} outlines the gravitational lensing setup, including the lens equation, deflection angle, and strong lensing coefficients for two different black holes solutions exhibiting DM halos in the background of the cloud of strings and quintessence field. Additionally, this section covers strong lensing observables, such as the position of the innermost image, image separation, brightness differences among relativistic images, and the time delay for supermassive black holes M87* and Sgr A*. In Section \ref{sec4}, we compare the strong lensing observables among the standard Schwarzschild black hole and black hole spacetime with a DM halo in two different models: the universal rotation curve (URC) and cold dark matter (CDM) halo with Navarro-Frenk-White (NFW). Finally, in Section \ref{sec5}, we summarize our findings and draw conclusions. Throughout the paper, we adopt units where the speed of light and the gravitational constant are set to 1 ($8\pi G = c = 1$); however, we restore these units in the tables for clarity."
\section{Black Hole Geometry and DM Halo}\label{sec2} 
In this section, we start with the spherically symmetric static spacetime to discuss the black hole geometry, which is described by the metric as,
\begin{equation}\label{1}
ds^2  =  - f(r) dt^2 + \frac{1}{f(r)} dr^2 + r^2 d\theta^2 + r^2 \sin^2\theta d\varphi^2,
\end{equation}
In the above equation, $f(r)$ defines the lapse function of black hole. For the current study, we shall consider the following lapse function \cite{mustafa2021radial, toledo2018black}:
\begin{equation}\label{2}
f(r)=\left(1-\alpha-\frac{2 M G_N}{c^2 r}-\frac{\gamma}{r^{3\omega_{q}+1}}\right)
\end{equation}
In the above equation, $\alpha$, $\gamma$, and $\omega_{q}$ are representing cloud of strings, quintessence, and quintessence field respectively with mass $M$, speed of light $c$, and the gravitational constant$G_N$. For the current analysis, we shall consider only one case, i.e., quintessence dark energy by fixing  $\omega_{q}=\frac{-2}{3}$ \cite{mustafa2021radial, toledo2018black}. Now, we will include the effect of two different kind of DM halos in the above black hole geometry, which are known as URC profile and CDM halo with NFW profile. 

\subsection{URC Profile}
In this subsection, we shall shortly discuss the well-known DM halo-like URC profile. URC matter distribution was described by \cite{salucci2000dark} and it is defined with the following expression
\begin{equation}\label{3}
\rho(r) = \frac{\rho_0 r_0^3}{(r+r_0)(r^2+r_0^2)}.
\end{equation}
In the above matter distribution, $r_0$ denotes the characteristic radius, and $\rho_0$ is the central density of the URC DM halo. According to the \cite{donato2009constant, salucci2019distribution} the best fit values within the recent observations of M87* galaxy for the parameters of the URC DM profile are $\rho_0 = 6.9\times 10^6 \text{M}_{\odot}/{\rm kpc}^{3}$ and $r_0 = 91.2\; {\rm kpc}$. For the Milky Way galaxy, both the involved parameters in URC DM halo are measured as $\rho_0 = 5.2 \times 10^7 \text{M}_{\odot}/{\rm kpc}^{3}$ and $r_0 = 7.8\; {\rm kpc}$ \cite{lin2019dark}. With this halo profile, the function $f(r)$ in the metric \eqref{2} is given by \cite{jusufi2019black}
\begin{widetext}
\begin{eqnarray}\label{4}
f(r) &=& e^{-\frac{2 \pi ^2 \rho _0 r_{0}^2 G_N}{c^2}} \left(\frac{r^2}{r_{0}^2}+1\right)^{-\frac{\left(1-\frac{r}{r_{0}}\right) \left(2 \pi  \rho _0 r_{0}^3 G_N\right)}{c^2 r}} \left(\frac{r}{r_{0}}+1\right)^{-\frac{\left(\frac{r}{r_{0}}+1\right) \left(4 \pi  \rho _0 r_{0}^3 G_N\right)}{c^2 r}} e^{\frac{4 \pi  \rho _0 r_{0}^3 G_N \tan ^{-1}\left(\frac{r \left(\frac{r}{r_{0}}+1\right)}{r_{0}}\right)}{c^2 r}}\nonumber\\&-&\frac{2 M G_N}{c^2 r}-\frac{\gamma}{r^{-1}}-\alpha 
\end{eqnarray}
\end{widetext}
Here $M = 6.5 \times 10^9~\text{M}_{\odot}$ for the M87* central black hole and $M = 4.3 \times 10^6~\text{M}_{\odot}$ for the Sgr A* black hole. The lapse function in the Eq. \eqref{4} is very complicated. In order to discuss strong gravitational lensing, we shall use the series solution just for the sack of simplicity, which is straightaway calculated as:
\begin{widetext}
\begin{eqnarray}\label{5}
f(r)&=&\left(1-2 \pi ^2 A B\right) \left(1-4 \pi  A B \left(\frac{B c^2 r}{M G_N}+1\right)\right) \left(4 \pi  A B \left(\frac{B c^2 r}{M G_N}+1\right)+1\right)\nonumber\\&\times& \left(\frac{2 \pi  A B^2 c^2 r \left(B c^2 r-M G_N\right)}{M^2 G_N^2}+1\right)-\frac{2 M G_N}{c^2 r}-\frac{\gamma}{r^{-1}} -\alpha,
\end{eqnarray}
\end{widetext}
where $A=\frac{\rho _0 r_{0}^{3}}{M}$ and $B=\frac{M G_N}{c^2 r_{0}}$. Now, the involved parameters $A$ and $B$ under the observational well-fitted values for M87* are calculated as:
$A=\frac{\rho _0 r_{0}^{3}}{M}=805.231;$
$B=\frac{M G_N}{c^2 r_{0}}=3.40611*10^{-9}$.
Further, for Sgr A* the parameters $A$ and $B$ are calculated as:
$A=\frac{\rho _0 r_{0}^{3}}{M}=5738.77;$
$B=\frac{M G_N}{c^2 r_{0}}=2.6346*10^{-11}.$
\subsection{The CDM Halo with NFW Profile}
The CDM halo model with NFW profile was calculated by using $N$-body simulations, which is also known as universal spherically averaged density profile \cite{navarro1997universal} and it is expressed as: 
\begin{equation}\label{6}
\rho(r) = \frac{\rho_0}{(r/r_0)(1+r/r_0)^2},
\end{equation}
In the above Eq. \eqref{5}, $\rho_0$ represents the density of the universe at the stage of collapsing of halo and $r_0$ denotes the characteristic radius. The recent observational data for the Milky Way galaxy \cite{lin2019dark}, the best-fit values for the involved parameters, i.e., $\rho_0$ and $r_0$ for CDM halo model with NFW are measured as $\rho_0 = 5.23 \times 10^7 \text{M}_{\odot}/{\rm kpc}^{3}$ and $r_0=8.1 \; {\rm kpc}$. For the other black hole model say M87* galaxy, the involved parameters are measured as $\rho_0 = 0.008 \times 10^{7.5}~\text{M}_{\odot}/ \text{kpc}^3$ (see \cite{oldham2016galaxy}) and ${r_0} = 130~\text{kpc}$ \cite{jusufi2019black}. Within scope of CDM halo profile, the function $f(r)$ in the metric by Eq. (\ref{2}) is provided by \cite{xu2018black} with the following form:
\begin{equation}\label{7}
f(r)=\left(1+\frac{r}{{r_0}}\right)^{-\frac{8 \pi G_N \rho_0 r_0^3}{c^2 r}} - \frac{2 G_N M}{c^2 r}-\frac{\gamma}{r^{-1}}-\alpha .
\end{equation}
Here in the above equation $M = 4.3 \times 10^6~\text{M}_{\odot}$ is the mass for Sgr A* black hole model and $M = 6.5 \times 10^9~\text{M}_{\odot}$ is the mass for M87 central black hole. Due to the involvement of the exponential function in the lapse function by Eq. \eqref{7} is very complicated for the discussion of strong gravitational lensing. Again, one can use the series solution for the lapse function of the black hole under the effect of CDM halo model with NFW profile, we have the following revised form of the lapse function:
\begin{eqnarray}\label{8}
f(r)&=&1-\frac{2M}{r}-\frac{\gamma}{r^{-1}} -\alpha +32 \pi ^2 A^2 B^2\nonumber\\&+&\frac{4 \pi  A B (B r+2 M)}{M},
\end{eqnarray}
where the values of parameters $A$ and $B$ for M87* is measured as: $A=\frac{\rho _0 r_{0}^{3}}{M}=85.508;$
$B=\frac{M G_N}{c^2 r_{0}}=2.38952*10^{-9}$. Further for the Sgr A* black model values of $A$ and $B$ are calculated as: $A=\frac{\rho _0 r_{0}^{3}}{M}=6463.81;$
$B=\frac{M G_N}{c^2 r_{0}}=2.537*10^{-11}$. 
\section{Strong Gravitational Lensing and It's Observable by Black Hole with DM Halos}\label{sec3} 
Here, we study the strong gravitational lensing by black holes with two different DM Halo models. Here, we investigate the deflection of photon rays in the equatorial plane ($\theta=\frac{\pi}{2}$) due to URC, and CDM halos models. To study the  deflection angle of photon rays in the equatorial plane ($\theta=\frac{\pi}{2}$), we write the Eq. \eqref{1} by the transformations, $t\rightarrow \frac{t}{2M}$, $r\rightarrow
\frac{r}{2M}$, $2M\gamma\rightarrow \gamma$ as:
 \begin{equation}\label{9}
d\bar{s}^2=-P(r)dt^2+ R(r) dr^2 +S(r) d\phi^2. 
\end{equation}
Now, by using the Eq. \eqref{9}, one can write 
the lapse function by Eq. \eqref{5} for the URC halo model  within the scope of Eq. \eqref{9} is expressed as:
\begin{widetext}
\begin{eqnarray}\label{10}
 P(r)=  \left(1-2 \pi ^2 A B\right) \left(4 \pi  A B^2 r (2 B r-1) +1\right) (4 \pi  A B (2 B r+1)+1) (1-4 \pi  A B (2 B r+1))-\frac{1}{r} -\gamma r-\alpha
\end{eqnarray}
\end{widetext}
Further, we have the following expression:
\begin{widetext}
\begin{eqnarray}\label{11}
 R(r)&=&\bigg[\left(1-2 \pi ^2 A B\right) \left(4 \pi  A B^2 r (2 B r-1) +1\right) (4 \pi  A B (2 B r+1)+1) (1-4 \pi  A B (2 B r+1))-\frac{1}{r} -\gamma r-\alpha \bigg]^{-1}
\end{eqnarray}
\end{widetext}
\begin{equation}\label{12}
    S(r)=r^2.
\end{equation}
The lapse function by Eq. \eqref{8} for the CDM with NFW profile model as:
\begin{eqnarray}\label{13}
    P(r)&=&1-\frac{1}{r}+32 \pi ^2 A^2 B^2+8 \pi  A B (B r+1)\nonumber\\&-&\gamma r -\alpha 
\end{eqnarray}
According to Eq. \eqref{1} and Eq. \eqref{9}, we have the following relation
\begin{eqnarray}\label{14}
    R(r)&=&\bigg[1-\frac{1}{r}+32 \pi ^2 A^2 B^2+8 \pi  A B (B r+1)\nonumber\\&-&\gamma r -\alpha \bigg]^{-1} 
\end{eqnarray}
\begin{equation}\label{15}
    S(r)=r^2,
\end{equation}
By utilizing Eq \eqref{9}, one can derive null geodesics with respect to the affine parameter $\tau$ using the following equations
   \begin{equation}\label{16}
   \dot{t}=\frac{dt}{d\tau}=\frac{E}{P(r)}
       \end{equation}

   \begin{equation}\label{17}
   \dot{\phi}= \frac{d\phi}{d\tau}=\frac{L}{r^2}
   \end{equation}

   \begin{equation}\label{18}
 \biggr( \frac{dr}{d\tau}\biggr)^2 = \dot{r}^2=E^2-\frac{L^2 }{r^2}P(r)
   \end{equation}
Here, the constants $E$ and $L$ represent the energy and angular momentum of the particle, respectively. The function $P(r)$ is defined by Eqs \eqref{10} and \eqref{13} for the cases of  URC and CDM halo respectively.
The Eq \eqref{18} can be written as
\begin{equation}\label{19}
   \left(\frac{dr}{d\tau}\right)^2+V_{eff} =E^2
\end{equation}
where the effective potential of a  photon is given by
\begin{equation}\label{20}
  V_{eff}=\frac{L^2 }{r^2}P(r)
\end{equation}
For the unstable circular photon orbit of radius $r_{ph}$, the conditions for the effective potentials  are$\frac{dV_{eff}}{dr}|_{r_{ph}}=0$ and
$\frac{d^2V_{eff}}{dr^2}|_{r_{ph}}<0$. Thus the radius of the photon sphere $r_{ph}$ is the largest real  root of the equation
\begin{equation}\label{21}
    2P(r_{ph})-r_{ph}P^{\prime}(r_{ph})=0
\end{equation}
Note that at $r=r_{ph}$, the condition  $\frac{d^2V_{eff}}{dr^2}|_{r_{ph}}<0$ is satisfied for the black hole spacetime \eqref{9}. Because these orbits are unstable against small perturbations, photons coming from infinity and approaching the black hole with a specific impact parameter denoted as "u," ultimately return to infinity after reaching their closest distance, $r_0$. When the particle reaches this minimum distance, $r_0$, closest to the central black hole, at which point $\frac{dr}{d\tau}=0$, we can express the ratio $\frac{L}{E}$ as a function of the impact parameter "u" in relation to the closest distance $r_0$ as \cite{Bozza:2002zj}
\begin{equation}\label{22}
   u=\frac{L}{E}=\frac{r_0}{\sqrt{P(r_0)}}
\end{equation}
The strong deflection angle experiences an infinite increase as $r_0 \rightarrow r_{ph}$ and remains finite only for When $r_0 > r_{ph}$. Thus the critical impact parameter, denoted as $u_{ph}$, is precisely defined by
 \begin{equation}\label{23}
u_{ph} =\frac{r_{ph}}{\sqrt{P(r_{ph})}}
 \end{equation}
For impact parameters less than $u_{ph}$, photons are drawn into the black hole, while for impact parameters greater than $u_{ph}$, photons approach their nearest distance to the black hole, denoted as $r_0$. However, when the impact parameter equals $u_{ph}$, photons follow an unstable circular orbit around the black hole, resulting in the formation of a photon sphere with a radius $r_{ph}$. The expression for the deflection angle in terms of the nearest approach distance $r_0$, within the context of the spacetime Eq \eqref{9}, can be described as follows: \cite{Virbhadra:2002ju,Claudel:2000yi}
 \begin{equation}\label{24}
\alpha_D(r_0)=I(r_0)- \pi
\end{equation}
where $r_0$ is the closest approach distance of photons trajectory and $I(r_0)$  is defined as
\begin{equation}\label{25}
I(r_0)=2 \int _{r_0}^\infty  \frac{d\phi}{dr} dr
\end{equation}
or,
\begin{equation}\label{26}
 I(r_0)=  \int_{r_0}^\infty \frac{2\sqrt{R(r)}dr}{\sqrt{S(r) \sqrt{ \frac{P(r_0)S(r)}{S(r)S(r_0)}-1} }}dr
 \end{equation}
In the strong field limit, the deflection angle $\alpha_D(r_0)$ is influenced by the relationship between $r_0$ and $r_{ph}$. Notably, when $r_0\approx r_{ph}$, it experiences an augmentation.
So, we define a new variable z as \cite{Bozza:2002zj,Chen:2009eu,Kumar:2021cyl}
\begin{equation}\label{27}
 z=1-\frac{r_0}{r}
\end{equation}
The total azimuthal angle, expressed in relation to these newly introduced variables, can be formulated as follows:
\begin{equation}\label{28}
 I(r_0)=\int_{0}^1 F(z,r_0)H(z,r_0)dz
\end{equation}
where
\begin{equation}\label{29}
F(z,r_0)= \frac{2(1-P(r_0))\sqrt{S(r_0)}}{S(r)A^\prime(r)}\sqrt{P(r)R(r)},
\end{equation}
  \begin{equation}\label{30}
 H(z,r_0)=\frac{\sqrt{S(r)}}{\sqrt{S(r)P(r_0)-P(r)S(r_0)}}
     \end{equation}
For all values of $z$ and  $ r_0 $ ,the function $ F(z,r_0 )$ becomes regular but the function  $H( z,r_0) $ diverges  at $ z=0 $ only. The integral \eqref{28} can be expressed as
 \begin{equation}\label{31}
     I(r_0) =  I^{D}(r_0)+I^{R}(r_0)
 \end{equation}
with  the regular part
 \begin{equation}\label{32}
 I^{R}(r_0)=\int_{0}^1 g(z,r_0) dz
\end{equation}
and the divergent part
\begin{equation}\label{33}
 I^{D}(r_0)=\int_{0}^1 F(0,r_{ph})H_0(z,r_{0}) dz
\end{equation}
where $g(z,r_0)=F(z,r_0)H(z,r_0) - F(0, r_{ph}) H_0(z,r_0) $. To
obtain the   divergence of the integrand in equation(22), one can expand  the  portion of   square root in $ R(z,r_0)$  as
  \begin{equation}\label{34}
      H_0(z,r_0)= \frac{1}{\sqrt{\eta(r_0) z +\zeta(r_0)  z^2 + \mathcal{O}(z^3) }}
  \end{equation}
where
\begin{equation}\label{35}
\begin{split}
\eta(r_0) =  \frac{(1 - P(r_0))}{P^\prime(r_0) S(r_0)} \bigg( P(r_0) S^\prime(r_0) \\
- P^\prime(r_0) S(r_0) \bigg)
\end{split}
\end{equation}
\begin{equation}\label{36}
 \begin{split}
&\zeta (r_0) = \frac{(1-P(r_0))^2}{{P^\prime }^3(r_0)
S^2(r_0)}\biggr(2S(r_0)S^\prime(r_0) {P^\prime(r_0)}^2 
+ (S(r_0) S^{\prime\prime}(r_0)\\
&-
2{S^\prime}^2(r_0))P(r_0)P^\prime(r_0) - S(r_0)S^\prime(r_0)P(r_0)P^{\prime\prime}(r_0)
 \biggr)\\
 \end{split}
\end{equation}
Here, the prime symbol represents differentiation with respect to the variable $r$. When $ r_0 \approx r_{ph}$, the coefficients $\eta(r_0)$ vanishes and the order of divergence is $z^{-1}$ which leads the integral divergence logarithemcally. For, $r_0 \approx r_{ph}$, the strong  deflection angle becomes
\cite{Tsukamoto:2016qro,Iyer:2006cn,Tsukamoto:2016jzh,Tsukamoto:2022uoz,Tsukamoto:2022tmm}
\begin{equation}\label{37}
\alpha(u)= -\bar{a}~ log\left(\frac{u}{u_{ph}}-1\right) +\bar{b} +\mathcal{O}((u -u_{ph})log(u -u_{ph}))
\end{equation}
where
\begin{equation}\label{38}
\bar{a}=\frac{F(0,r_{ph})}{2\sqrt{\zeta(r_{ph})}}
=\sqrt{\frac{2P(r_{ph}) R(r_{ph})}{P(r_c)S^{\prime\prime}(r_{ph})-P^{\prime\prime}(r_{ph}) S(r_{ph})}}
\end{equation}
and
\begin{equation}\label{39}
\bar{b}=-\pi + a_R + \bar{a} log \biggr({\frac{4\zeta(r_{ph})P(r_{ph})}{u_{ph} |P(r_{ph})| ( 2u_{ph} P(r_{ph}))}}\biggr),
\end{equation}
$ a_R=I^{R}(r_{ph})= \int_{0}^{1} g(z,r_{ph} ) dz  $ which is numerically obtained.\\\\\
Now, we discuss some strong lensing observables quantities  With the help of strong lensing coefficients $\mathit{u_{ph}}$,$\mathit{\bar{a}}$,$\mathit{\bar{b}}$ and strong deflection angle $\alpha_D$. In this context, we examine a scenario in which the source, the black hole acting as a lens, and the receiver are perfectly alignment, and both the receiver and source are positioned at a very far from the black hole (lens). Thus one can express the lens equation as
\cite{Bozza:2001xd}
\begin{equation}\label{40}
 \beta=\tilde{\alpha}-\frac{d_{ls}}{d_{os}}\Delta \alpha_{n}
\end{equation}
Here, we use the variables $\beta$ and $\tilde{\alpha}$ to represent the angular positions of the image and the source, respectively, relative to the optical axis. Additionally, we denote the distances between the lens and the source as $d_{ls}$, between the receiver and the lens as $d_{ol}$, and between the receiver and the source as $d_{os}$, respectively such that $d_{os} = d_{ol} + d_{ls}$. Here, we define $\Delta\alpha_{n}$ as the angular offset, where $\Delta\alpha_{n} = \tilde{\alpha} - 2n\pi$ and the variable $n$ signifies the number of complete loops made by the light ray. Using the Eqs. \eqref{37} and \eqref{40}, the angular separation between  the $n^{th}$ relativistic  and the black hole (lens) can be
written as
\begin{equation}\label{41}
\theta _n =  \theta^0 _n - \frac{u_{ph} e_n  d_{os}(\theta_n^0-\beta)}{\bar{a}d_{ls}d_{ol}}
\end{equation}
where $$ e_n=e^{\frac{\bar{b}-2n\pi}{\bar{a}}},$$
$$\theta^0_n=\frac{ u_{ph}(1+e_n)}{d_{ol}},$$ $\theta^0_n$ is  the
image position for $\alpha=2n\pi$. The magnification  for the  $n$-th relativistic image, is defined as \cite{Bozza:2002zj}
\begin{equation}\label{42}
\mu_n=\biggr(\frac{\beta}{\tilde{\alpha}}\frac{d\beta}{d\alpha}\biggr)^{-1}\biggr|_{\theta_0}=\frac{e_n u^2_{ph}(1+e_n)d_{os}}{\bar{a}\beta d_{ls}d^2_{ol}}
\end{equation}
Clearly, the initial (i.e first) relativistic image shines the brightest, and the magnification decreases exponentially as we move to higher image orders denoted by $n$. Importantly, as  $\beta \rightarrow 0$, Eq \eqref{42} exhibits divergence. Consequently, perfect alignment significantly enhances the possibility of detecting the images. If we designate $\theta_n$ as the asymptotic position where a cluster of images converges as $n$ approaches infinity, then the brightest image, specifically the outermost image denoted as $\theta_1$, can be individually distinguished. All other images become tightly grouped together at the position $\theta_{\infty}$, which represents the asymptotic location of the set of relativistic images as in the limit $n \rightarrow \infty $. By utilizing the deflection angle as described in Eq \eqref{37} and the lens equation provided in Eq \eqref{40}, we have derived three observable quantities 
such as the angular position of the set of images
$\theta_{\infty}$, angular separation $S$  between the outermost
image and remaining the inner set of images, and the flux ratio between the outermost relativistic image and the remaining inner set of relativistic images can be expressed as
\cite{Kumar:2022fqo,Bozza:2002zj}
\begin{equation}\label{43}
\theta_{\infty}=\frac{u_{ph}}{d_{ol}}
\end{equation}
\begin{equation}\label{44}
S= \theta_1-\theta_{\infty}\approx\theta_{\infty}e^\frac{(\bar{b} -2\pi)}{\bar{a}}
\end{equation}
\begin{equation}\label{45}
r_{mag}=\frac{\mu_1}{\Sigma^\infty_{n=2}\mu_{n}}\approx \frac{5\pi}{\bar{a}log(10)}
\end{equation}
If the observables quantities $\theta_{\infty}$,$S$,  and $r_{mag}$ are obtained from the observation, the strong lensing coefficients $\bar{a}$,$\bar{b}$ and the minimum impact parameter $u_{ph}$  can be obtained easily by the Eqs \eqref{43}, \eqref{44} and \eqref{45}, and also, this value can be compared to the predictions from theoretical models. Thus, one can be identified the characterization of the black hole in the presence of a DM halo. Another important observable quantity is the time delays $\Delta T_{2,1}$  between two different relativistic images. Time travel along the photon path varies for different relativistic images, resulting in time differences between these images. This time delay serves as a crucial observable, derived from the disparity in image formation times. By distinguishing the time signals of the first and second images during observations, it is possible to calculate the time delay between these two signals \cite{Bozza:2003cp}. As a photon travels from the source to an observer, the time it takes to traverse a path around the black hole can be determined using the formula provided by \cite{Bozza:2003cp}.
\begin{equation}\label{46}
\tilde{T}=\tilde{a}log\biggr(\frac{u}{u_{ph}}-1\biggr)+\tilde{b}+\mathcal{O}(u-u_{ph})
\end{equation}
Employing the Eq \eqref{46} makes it feasible to calculate the time disparity between the two relativistic images. In the context of spherically symmetric spacetime, the time delay between two relativistic images ( first and second ) positioned on the same side of the black hole can be expressed as follows \cite{Bozza:2003cp} :
\begin{equation}\label{47}
\Delta T_{2,1}=2\pi u_{ph}=2\pi D_{ol} \theta_{\infty}
\end{equation}
Using the above formula \eqref{47}, it is beneficial to investigate the time delay associated with various black holes positioned at the center of neighboring galaxies. In this paper, we investigate the time delay associated with two supermassive black holes $M87^*$ and $SgrA^{*}$ in the context of black hole with two significant URC, CDM halo.
\subsection{Strong Lensing Observables for URC Halo}
\begin{figure*}[htbp]
\centering
\includegraphics[width=.45\textwidth]{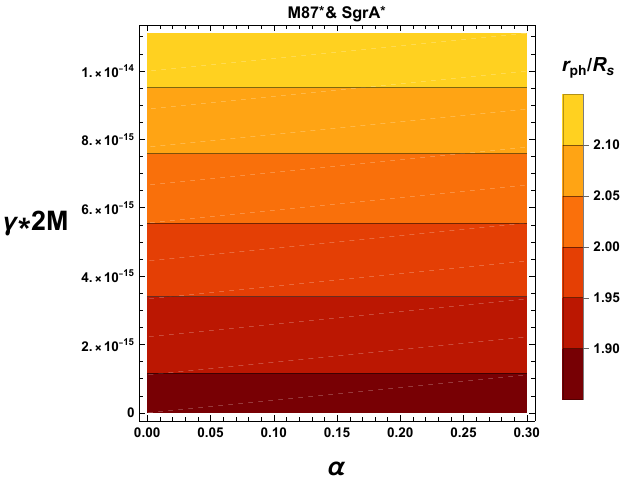}(a)\label{fig1a}
\qquad
\includegraphics[width=.45\textwidth]{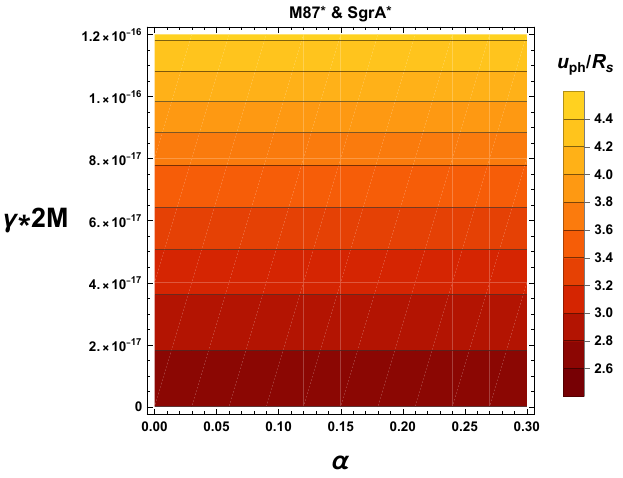}(b)\label{fig1a}
\caption{The behavior of the photon sphere radius $\mathit{r_{ph}}$(left panel) and the minimum impact parameter $\mathit{u_{ph}}$ (right panel) as a function of both the parameters $\alpha$ and $\gamma \cdot 2M$ for the URC model. It's worth noting that the photon sphere radius $\mathit{r_{ph}}$  for the cases of $M87^{*}$ and $Sgr A^{*}$exhibit nearly identical behavior and also the minimum impact parameter $\mathit{u_{ph}}$ for the cases of $M87^{*}$ and $Sgr A^{*}$ exhibit nearly identical behavior.
}\label{fig:1}
\end{figure*}

\begin{figure*}[htbp]
\centering
\includegraphics[width=.45\textwidth]{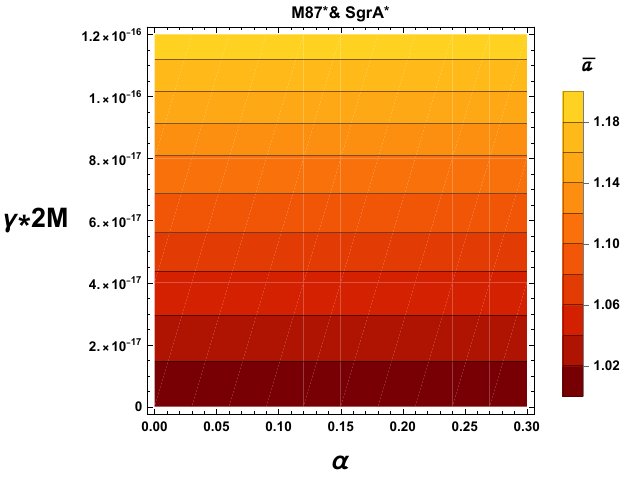}(a)
\qquad
\includegraphics[width=.45\textwidth]{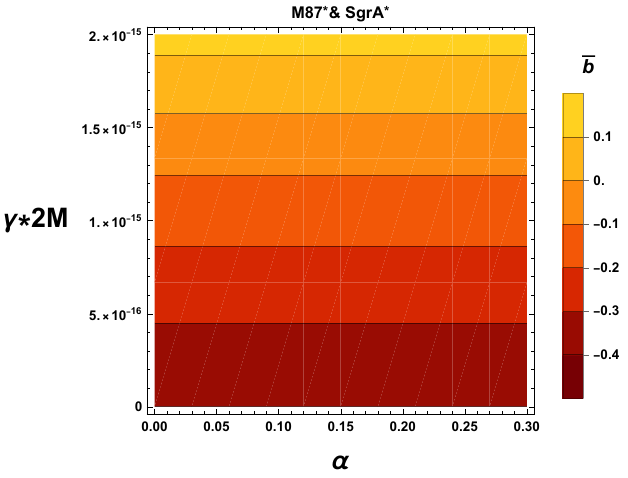}(b)
\caption{The behaviour of the deflection limit coefficients $\mathit{\bar{a}}$(left panel) and  $\mathit{\bar{b}}$(right panel)  as a function of both the parameters $\alpha$ and $\gamma \cdot 2M$for the URC model . It's worth noting that the deflection limit coefficients $\mathit{\bar{a}}$  for the cases of $M87^{*}$ and $Sgr A^{*}$exhibit nearly identical behavior and also the deflection limit coefficients $\mathit{\bar{b}}$ for the cases of $M87^{*}$ and $Sgr A^{*}$exhibit nearly identical behavior.}\label{fig:2}
\end{figure*}

\begin{figure*}[htbp]
\centering
\includegraphics[width=.45\textwidth]{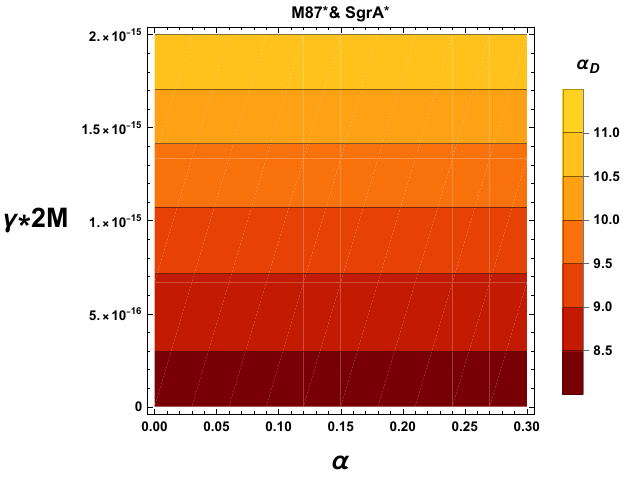}
\caption{The behaviour of the strong  deflection angle $\mathit{\alpha_{D}}$ as a function of both the parameters $\alpha$ and $\gamma \cdot 2M$ for the URC model.It's worth noting that the deflection limit coefficients $\mathit{\bar{a}}$  for the cases of $M87^{*}$ and $Sgr A^{*}$exhibit nearly identical behavior.}\label{fig:3}
\end{figure*}

\begin{figure*}[htbp]
\centering
\includegraphics[width=.45\textwidth]{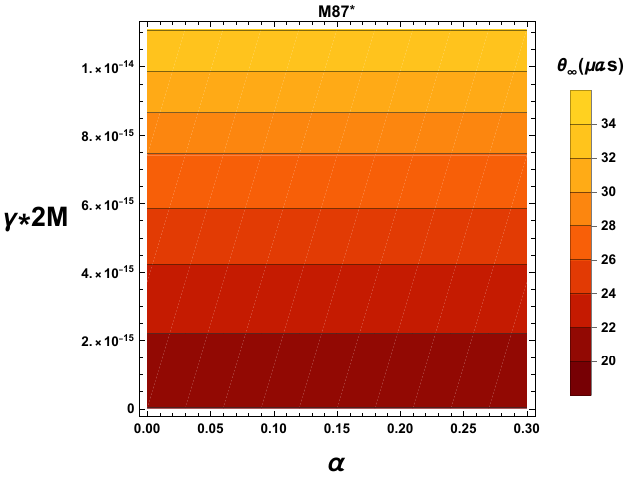}(a)
\qquad
\includegraphics[width=.45\textwidth]{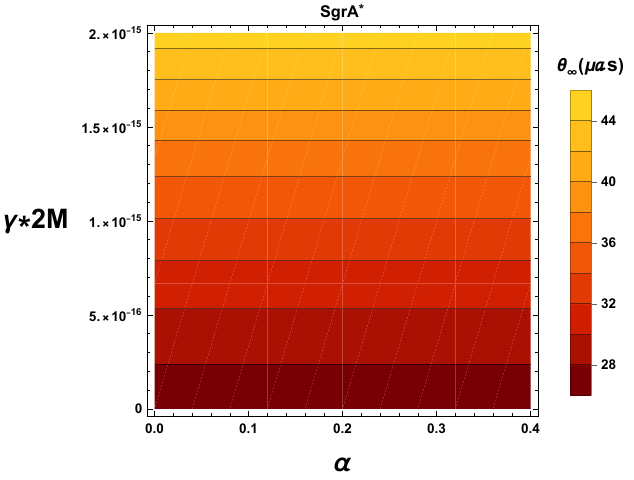}(b)
\caption{The behaviour of the observable quantity angular position $\mathit{\theta_{\infty}}$ for $M87^{*}$ (left panel) and  for $Sgr A^*$(right panel) as a function of both the parameters $\alpha$ and $\gamma \cdot 2M$ for the URC model.}\label{fig:4}
\end{figure*}

\begin{figure*}[htbp]
\centering
\includegraphics[width=.45\textwidth]{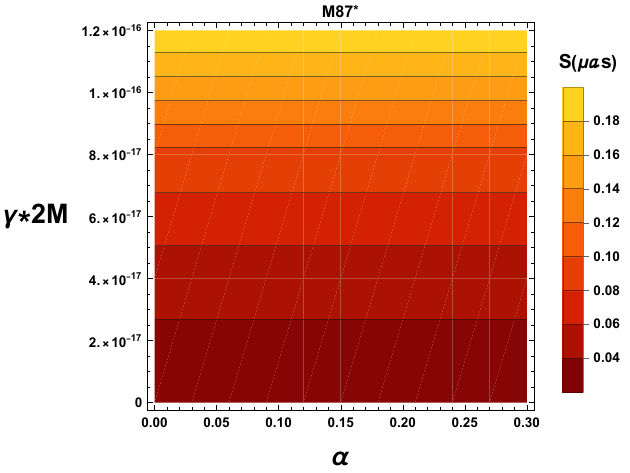}(a)
\qquad
\includegraphics[width=.45\textwidth]{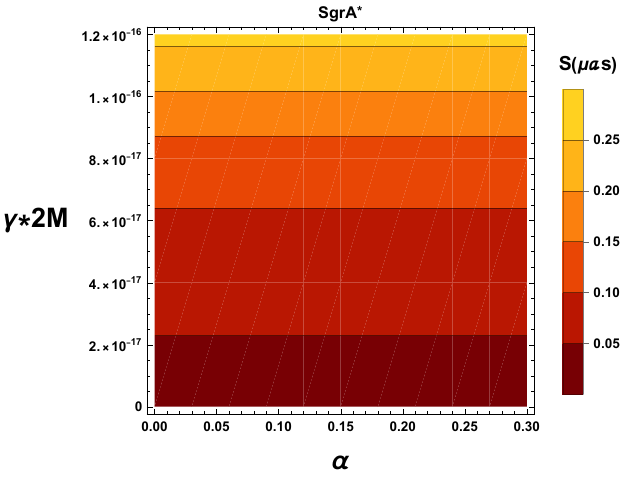}(b)
\caption{The behaviour of the observable quantity angular separation $S$ for $M87^{*}$ (left panel) and  for $Sgr A^*$(right panel) as a function of both the parameters $\alpha$ and $\gamma \cdot 2M$ for the URC model.}\label{fig:5}
\end{figure*}

\begin{figure*}[htbp]
\centering
\includegraphics[width=.45\textwidth]{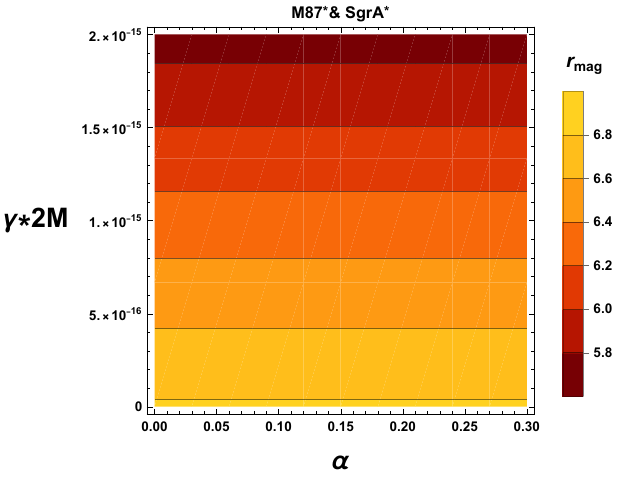}
\caption{The behaviour of the relative magnification $\mathit{r_{mag}}$ as a function of both the parameters $\alpha$ and $\gamma \cdot 2M$ for the URC model.It's worth noting that the relative magnification $\mathit{r_{mag}}$  for the cases of $M87^{*}$ and $Sgr A^{*}$exhibit nearly identical behavior.}\label{fig:6}
\end{figure*}

\begin{figure*}[htbp]
\centering
\includegraphics[width=.45\textwidth]{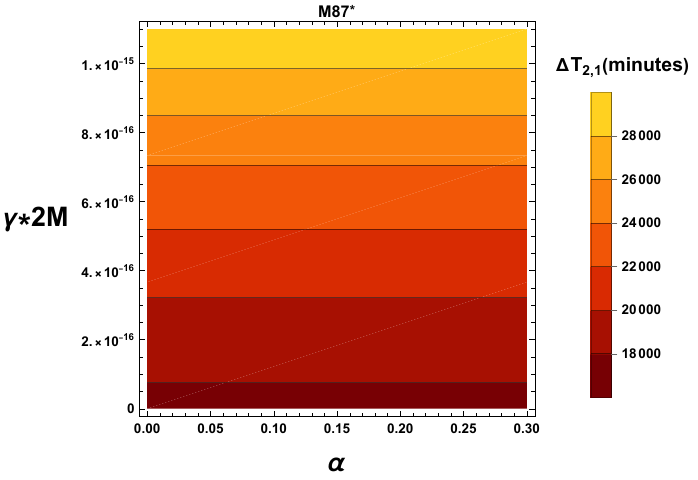}(a)
\qquad
\includegraphics[width=.45\textwidth]{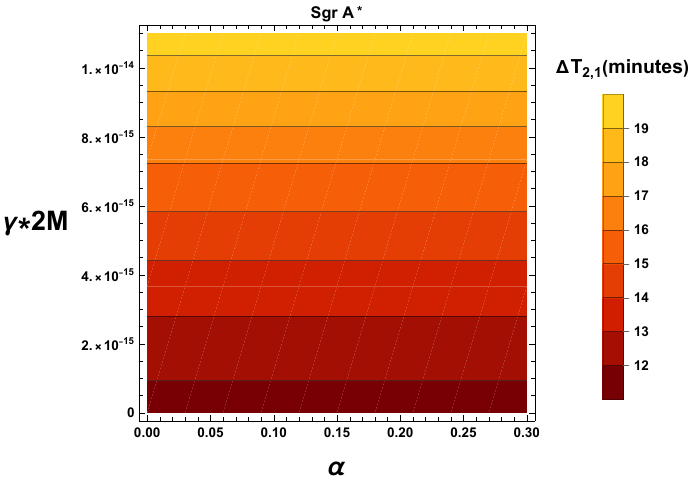}(b)
\caption{The behaviour of the time delays $\Delta T_{2,1}$ between two relativistic images for $M87^{*}$ (left panel) and  for $Sgr A^*$(right panel) as a function of both the parameters $\alpha$ and $\gamma \cdot 2M$ for the URC model.Time delays $\Delta T_{2,1}$ are measured in minutes .}
\label{fig:7}
\end{figure*}

Here, we discuss the strong gravitational lensing and its various observable with URC DM halo in the presence of cloud string and quintessence. The behavior of the photon sphere radius $\mathit{r_{ph}}$ in Fig.\ref{fig:1} a and the minimum impact parameter $\mathit{u_{ph}}$ in Fig.\ref{fig:1} b are described as a function of both the parameters $\alpha$ and $\gamma \cdot 2M$ for the case of URC model. From Fig.\ref{fig:1} a, it is observed that the photon sphere radius $\mathit{r_{ph}}$ increases with the parameters $\alpha$ for the fixed value of $\gamma \cdot 2M$ and also increases with the parameters $\gamma \cdot 2M$ for the fixed value of $\alpha$. From Fig.\ref{fig:1} b and Tables \ref{table:1} \& \ref{table:2}, it is observed that the minimum impact parameter $\mathit{u_{ph}}$  increases with the parameters $\alpha$ for the fixed value of $\gamma \cdot 2M$ and also increases with the parameters $\gamma \cdot2M$ for the fixed value of $\alpha$. It also observes that when $A=0$ and $B=0$, $\alpha=0$ and $\gamma \cdot 2M=0$ the photon sphere radius $r_{ph}=1.5$; and the minimum impact parameter $u_{ph}=2.59808$, corresponds to the case of Schwarzschild black hole \cite{Bozza:2002zj}.\\\\
The behavior of deflection limit coefficients $\mathit{\bar{a}}$ in Fig.\ref{fig:2} a and the deflection limit coefficients $\mathit{\bar{b}}$ in Fig.\ref{fig:2} b are described as a function of both the parameters $\alpha$ and $\gamma \cdot 2M$ for the case of URC model. From Fig.\ref{fig:2} a and Tables \ref{table:1} \& \ref{table:2}, it is observed that deflection limit coefficients $\mathit{\bar{a}}$ increases with the parameters $\alpha$ for the fixed value of $\gamma \cdot 2M$ and also increase with the parameters $\gamma \cdot 2M$ for the fixed value of $\alpha$. From Fig.\ref{fig:2} b and Tables \ref{table:1} \&\ref{table:2}, it is observed that deflection limit coefficients $\mathit{\bar{b}}$ increase with the parameters $\alpha$ for the fixed value of $\gamma \cdot 2M$ and also increase with the parameter $\gamma \cdot 2M$ for the fixed value of $\alpha$. It also observe that when $A=0$ and $B=0$,$\alpha=0$ and $\gamma \cdot 2M=0$ the the deflection limit coefficients $\mathit{\bar{a}}=1$; and the deflection limit coefficients $\mathit{\bar{b}}=-0.400023$ , corresponds to the case of Schwarzschild black hole \cite{Bozza:2002zj}.\\\\
The behavior of the deflection angle in Fig.\ref{fig:3} is described as a function of both the parameters $\alpha$ and $\gamma \cdot 2M$ for the case of the URC model. It is observed that the deflection angle $\alpha_D$ increases with the parameters $\alpha$ for the fixed value of $\gamma \cdot 2M$ and also increases with the parameters $\gamma \cdot 2M$ for the fixed value of $\alpha$. Moreover the  deflection angle $\alpha_D$ for the cases of $M87^{*}$ and $Sgr A^{*}$ exhibit nearly identical behavior. It's worth noting that the photon sphere radius $\mathit{r_{ph}}$, minimum impact parameter $\mathit{u_{ph}}$, deflection limit coefficients $\mathit{\bar{a}}$, deflection limit coefficients $\mathit{\bar{b}}$ and the strong deflection angle $\mathit{\alpha_D}$ for the cases of $M87^{*}$ and $Sgr A^{*}$ exhibit nearly identical behavior for the  URC model.\\\\
The behavior of observable quantities angular image position$ \theta_{\infty}$ in Fig.\ref{fig:4}, image separation $S$ in Fig.\ref{fig:5}, relative magnification $r_{mag}$ in Fig.\ref{fig:6} of the relativistic images and the time delays $\Delta T_{2,1}$  between two different relativistic images in Fig.\ref{fig:7} are described as a function of both the parameters $\alpha$ and $\gamma \cdot 2M$ for the case of URC model. From Figs.\ref{fig:4} \& \ref{fig:5}, it is observed that observable quantities angular image position$ \theta_{\infty}$ and image separation $S$  increases with the parameters $\alpha$ for the fixed value of $\gamma \cdot 2M$ and also increase with the parameters $\gamma \cdot 2M$ for the fixed value of $\alpha$; but in Fig. \ref{fig:6}, the observable quantity relative magnification $r_{mag}$ decreases with the parameters $\alpha$ for the fixed value of $\gamma \cdot 2M$ and also decreases with the parameters $\gamma \cdot 2M$ for the fixed value of $\alpha$ for both the cases of $M87^{*}$ and $Sgr A^{*}$ supermassive black hole (cf.Table.\ref{table:3}). From Fig.\ref{fig:7} a \& b  and Table.\ref{table:3}, it is observed that the time delays $\Delta T_{2,1}$  between two different relativistic images increases with the parameters $\alpha$ for the fixed value of $\gamma \cdot 2M$ and also increases with the parameters $\gamma \cdot 2M$ for the fixed value of $\alpha$ in the context of URC DM halo . 
\begin{table*}
\begin{tabular}{p{2.5cm}  p{3cm} p{2.5cm} p{4cm} p{2cm} }
\hline
\hline
\multicolumn{5}{c}{Strong Lensing Coefficients }\\
$\gamma \cdot 2M$ & $\alpha$& $\bar{a}$  & $\bar{b} $ &$ u_{ph}/R_{s}$\\
\hline
Schwarzschild&Black Hole&  1.00&-0.40023&2.59808\\
\hline
0&0&  1.00003&-0.400156&2.59829\\
\hline
&0.1& 1.05412&-0.251856&3.04318\\
&0.2&  1.11807&-0.0765524&3.63129\\
&0.3&  1.19527&0.135089&4.43665\\
&0.4&  1.29105&0.397651&5.59093\\
$1.2\times 10^{-16}$&0.5&  1.41429&0.735489&7.34966\\
&0.6&  1.58125&1.19318&10.2719\\
&0.7&  1.82591&1.86388&15.8157\\
&0.8&  2.23637&2.98911&29.0592\\
&0.9&  3.16313&5.5297&82.2251\\
\hline
\hline
\end{tabular}
\caption{Estimation  of strong lensing coefficients with the different value of black hole parameters  $\alpha$ and $\gamma*2M=1.2\times 10^{-16}$ for the case of  URC  for  $M87*$ black hole  where $A=805.231$; $B=3.40611*10^{-9} $ and $\omega=-\frac{2}{3}$.\label{table:1}}
\end{table*}
\begin{table*}
\begin{tabular}{p{2.5cm}  p{3cm} p{2.5cm} p{4cm} p{2cm} }
\hline
\hline
\multicolumn{5}{c}{Strong Lensing Coefficients }\\
$\gamma \cdot 2M$ & $\alpha$& $\bar{a}$  & $\bar{b} $ &$ u_{ph}/R_{sh}$\\
\hline
Schwarzschild&Black Hole&  1.00&-0.40023&2.59808\\
\hline
0&0&  1&-0.4002226&2.59809\\
\hline
&0.1& 1.05409&-0.251938&3.04292\\
&0.2&  1.11804&-0.0766504&3.63094\\
&0.3&  1.19523&0.134969&4.43616\\
&0.4&  1.291&0.3975&5.59021\\
$1.2\times 10^{-16}$&0.5&  1.41422&0.735291&7.34854\\
&0.6&  1.58114&1.1929&10.2699\\
&0.7&  1.82575&1.86345&15.8116\\
&0.8&  2.23608&2.98833&29.048\\
&0.9&  3.16232&5.52749&82.1621\\
\hline
\hline
\end{tabular}
\caption{Estimation  of strong lensing coefficients with the different value of black hole parameters  $\alpha$ and $\gamma*2M=1.2\times 10^{-16}$ for the case of  URC   for the case of $Sgr A^{*}$ black hole  where $A=5738.77,B=2.6346*10^{-11}$and $\omega=-\frac{2}{3}$.\label{table:2}}
\end{table*}

\begin{table*}
\begin{center}
\begin{tabular}{cc|cc|cc|cc|cc}
\hline
\hline
{parameters} && {$ M87^*$}& &{$ Sgr A^*$ }&&{$ M87^*$}&{$ Sgr A^*$}&$ M87^*$  &$Sgr A^*$
\\
$\gamma \cdot 2M$ & $\alpha$ & $ \theta_{\infty} (\mu as)$&$S(\mu as)$&
$\theta_{\infty} (\mu as)$&$S(\mu as)$ &$r_{mag}(magnitude)$&$r_{mag}(magnitude)$& $\Delta T_{2,1}(min)$ & $\Delta T_{2,1}(min)$ \\
\hline
\hline
Schwarzschild & black hole&19.9633&0.024984&26.3315&0.0329538&6.82188&6.82188&$17378.8$&11.4973\\
\hline
{0} & 0&19.9633&0.0249923&26.3827&0.0330183&6.8217&6.82187&$17378.02$&11.4974\\
\hline
 & 0.1&23.3833&0.0474787&30.8998&0.0627246&6.47161&6.47179&$20356.2$&13.4659\\
 & 0.2&27.9023&0.094475&36.8701&0.124809&6.10147&6.10167&$24290.1$&16.0681\\
 & 0.3&34.0906&0.198961&45.0478&0.262834&5.70738&5.70758&$29677.3$&19.6315\\
 & 0.4&42.9599&0.450029&56.7667&0.594479&5.28397&5.28419&$37398.4$&24.7385\\
$1.2\times 10^{-16}$ & 0.5&56.4737&1.11759&74.6219&1.47623&4.82354&4.82378&$49162.8$&32.5197\\
 & 0.6&78.9277&3.154676&104.288&4.16945&4.31424&4.31452&$68709.8$&45.4478\\
 & 0.7&121.525&10.8026&160.562&14.2663&3.73616&3.73648&$105793$&69.9717\\
 & 0.8&223.286&51.1877&294.973&67.5852&3.05043&3.05082&$194380$&128.547\\
 & 0.9&631.806&497.888&834.328&656.983&2.15668&2.15724&$550014$&363.594\\
\hline
\end{tabular}
\end{center}
\caption{Estimation of strong lensing observables for URC case for the supermassive black holes $ M87^*$, $ Sgr A^*$,  with the different value of black hole parameters  $\alpha$; and $q=1.2\times 10^{-16}$.The observable quantity $r_{mag}$ does not depend on the mass or distance of the black hole from the observer.\label{table:3}}
\end{table*}
\subsection{Strong lensing observables for  CDM Halo with NFW}
\begin{figure*}[htbp]
\centering
\includegraphics[width=.45\textwidth]{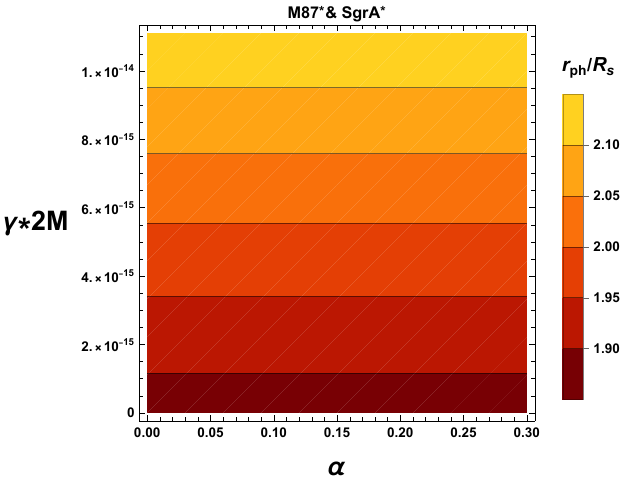}(a)
\qquad
\includegraphics[width=.45\textwidth]{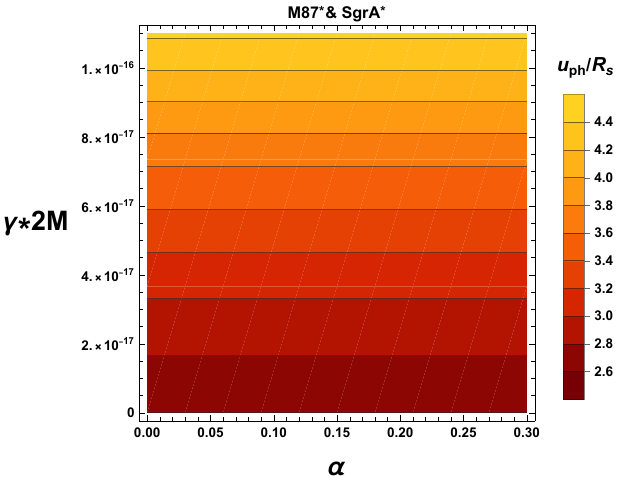}(b)
\caption{The behaviour of the photon sphere radius $\mathit{r_{ph}}$(left panel) and the minimum impact parameter $\mathit{u_{ph}}$(right panel) as a function of both the parameters $\alpha$ and $\gamma \cdot 2M$for the CDM model . It's worth noting that the photon sphere radius $\mathit{r_{ph}}$  for the cases of $M87^{*}$ and $Sgr A^{*}$exhibit nearly identical behavior and also the minimum impact parameter $\mathit{u_{ph}}$ for the cases of $M87^{*}$ and $Sgr A^{*}$exhibit nearly identical behavior.}\label{fig:8}
\end{figure*}

\begin{figure*}[htbp]
\centering
\includegraphics[width=.45\textwidth]{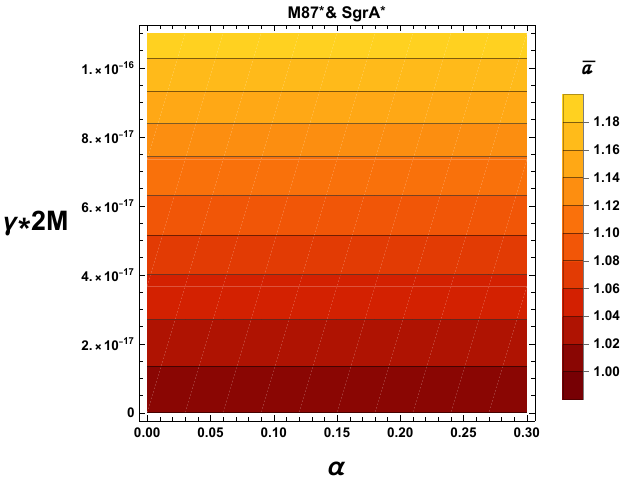}(a)
\qquad
\includegraphics[width=.45\textwidth]{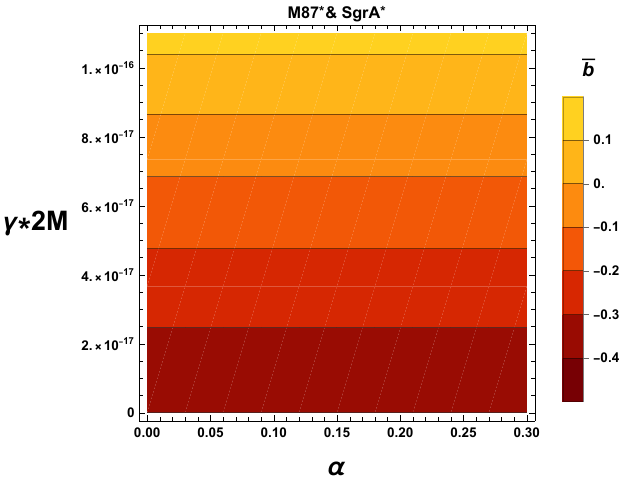}(b)
\caption{The behaviour of the deflection limit coefficients $\mathit{\bar{a}}$(left panel) and  $\mathit{\bar{b}}$(right panel)  as a function of both the parameters $\alpha$ and $\gamma \cdot 2M$ for the CDM model. It's worth noting that the deflection limit coefficients $\mathit{\bar{a}}$  for the cases of $M87^{*}$ and $Sgr A^{*}$exhibit nearly identical behavior and also the deflection limit coefficients $\mathit{\bar{b}}$ for the cases of $M87^{*}$ and $Sgr A^{*}$exhibit nearly identical behavior.}\label{fig:9}
\end{figure*}

\begin{figure*}[htbp]
\centering
\includegraphics[width=.45\textwidth]{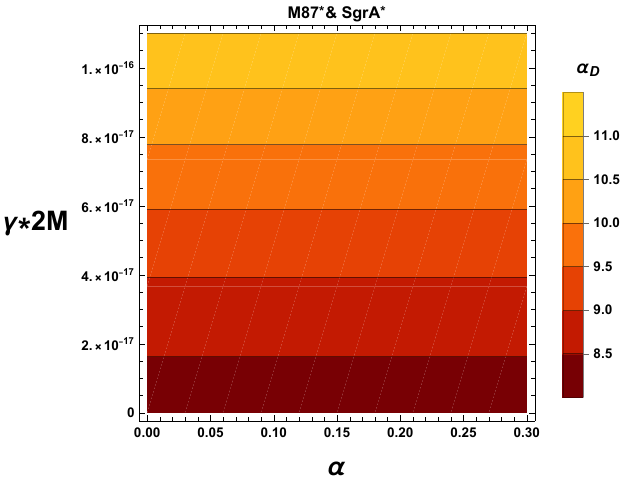}
\caption{The behaviour of the strong  deflection angle $\mathit{\alpha_{D}}$ as a function of both the parameters $\alpha$ and $\gamma \cdot 2M$ for the CDM model.It's worth noting that the deflection limit coefficients $\mathit{\bar{a}}$  for the cases of $M87^{*}$ and $Sgr A^{*}$exhibit nearly identical behavior.}\label{fig:10}
\end{figure*}

\begin{figure*}[htbp]
\centering
\includegraphics[width=.45\textwidth]{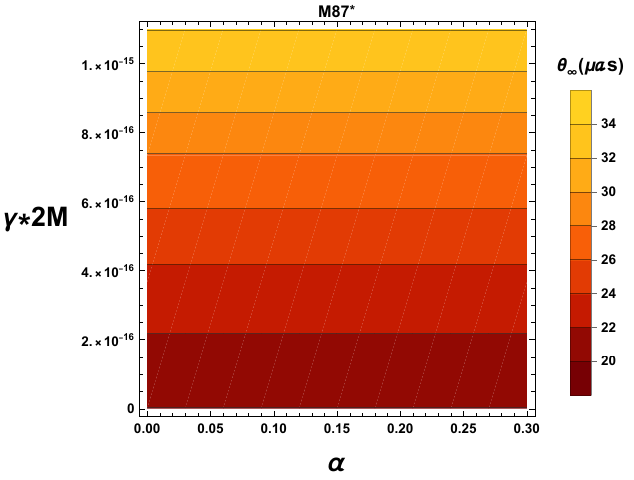}(a)
\qquad
\includegraphics[width=.45\textwidth]{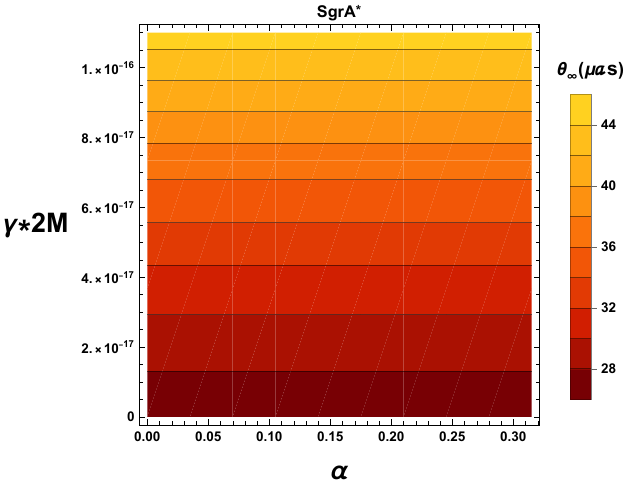}(b)
\caption{The behaviour of the observable quantity angular position $\mathit{\theta_{\infty}}$ for $M87^{*}$ (left panel) and  for $Sgr A^*$(right panel) as a function of both the parameters $\alpha$ and $\gamma \cdot 2M$ for the CDM model .}\label{fig:11}
\end{figure*}

\begin{figure*}[htbp]
\centering
\includegraphics[width=.45\textwidth]{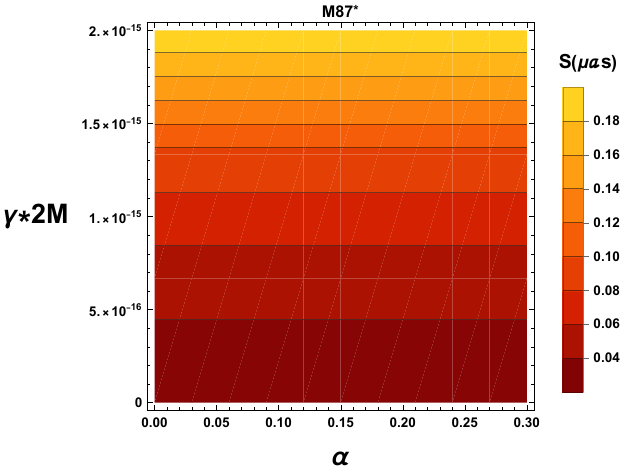}(a)
\qquad
\includegraphics[width=.45\textwidth]{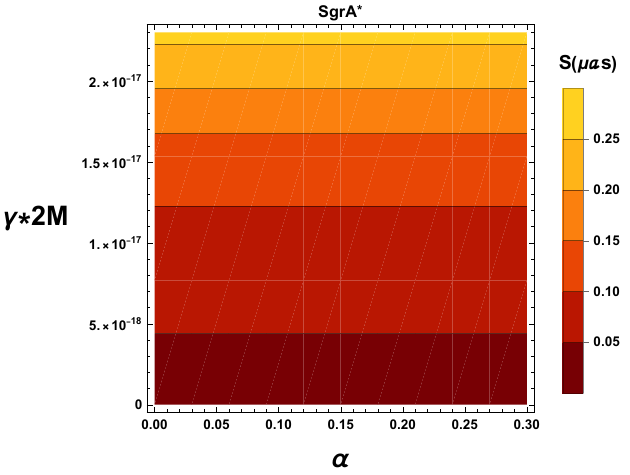}(b)
\caption{The behaviour of the observable quantity angular separation $S$ for $M87^{*}$ (left panel) and  for $Sgr A^*$(right panel) as a function of both the parameters $\alpha$ and $\gamma \cdot 2M$for the CDM model.}\label{fig:12}
\end{figure*}

\begin{figure*}[htbp]
\centering
\includegraphics[width=.45\textwidth]{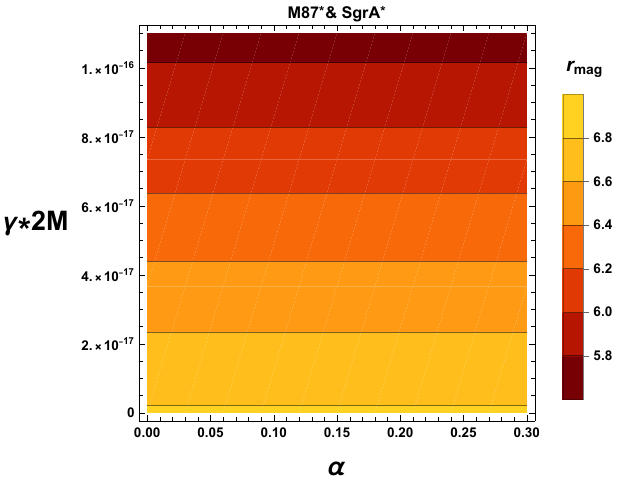}
\caption{The behaviour of the relative magnification $\mathit{r_{mag}}$ as a function of both the parameters $\alpha$ and $\gamma \cdot 2M$for the CDM model.It's worth noting that the relative magnification $\mathit{r_{mag}}$  for the cases of $M87^{*}$ and $Sgr A^{*}$exhibit nearly identical behavior.}\label{fig:13}
\end{figure*}

\begin{figure*}[htbp]
\centering
\includegraphics[width=.45\textwidth]{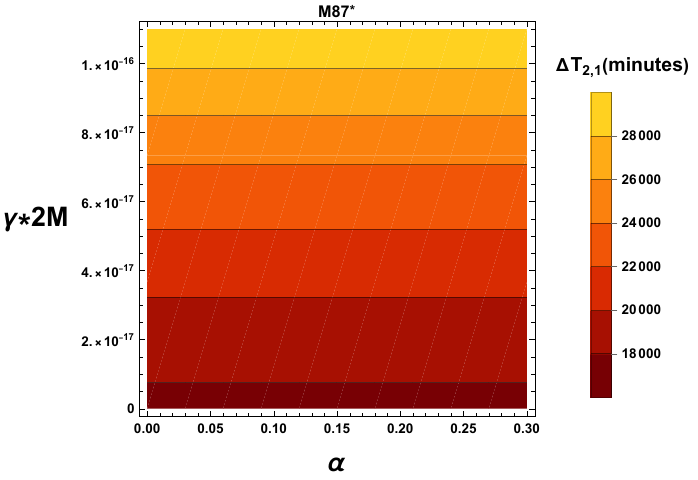}(a)
\qquad
\includegraphics[width=.45\textwidth]{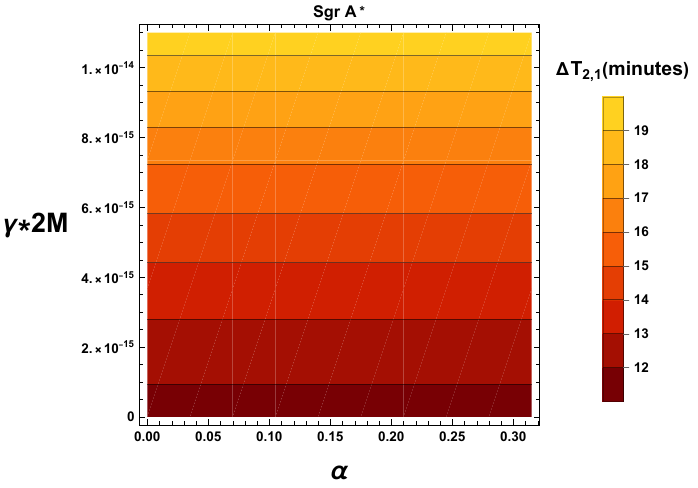}(b)
\caption{The behaviour of the time delays $\Delta T_{2,1}$ between two relativistic images for $M87^{*}$ (left panel) and  for $Sgr A^*$(right panel) as a function of both the parameters $\alpha$ and $\gamma \cdot 2M$ for the CDM model.Time delays $\Delta T_{2,1}$ are measured in minutes.}\label{fig:14}
\end{figure*}

Here, we discuss the strong gravitational lensing and its various observable with CDM halo in the presence of Cloud string and quintessence.
\begin{table*}
\begin{tabular}{p{2.cm}  p{2cm} p{2.5cm} p{4cm} p{2cm} }
\hline
\hline
\multicolumn{5}{c}{Strong Lensing Coefficients }\\
$\gamma*2M $& $\alpha$& $\bar{a}$  & $\bar{b} $ &$ u_{ph}/R_{sh}$\\
\hline
Schwarzschild&Black Hole&  1.00&-0.40023&2.59808\\
\hline
0&0&  0.999997&-0.400237&2.59806\\
\hline
&0.1& 1.05409&-0.251951&3.04288\\
&0.2&  1.11803&-0.0766659&3.63089\\
&0.3&  1.19522&0.13495&4.43609\\
&0.4&  1.1.29099&0.397476&5.5901\\
$5\times 10^{-17}$&0.5&  1.41421&0.73526&7.34836\\
&0.6&  1.58113&1.19285&10.2696\\
&0.7&  1.82573&1.86338&15.811\\
&0.8&  2.23604&2.9882&29.0463\\
&0.9&  3.1622&5.52713&82.1521\\
\hline
\hline
\end{tabular}
\caption{Estimation  of strong lensing coefficients with the different value of black hole parameters  $\alpha$ and $q=5\times 10^{-17}$   for the case CDM of $M87*$ black hole  where $A=85.508$;
$B=2.38952*10^{-9}$,  $\omega=-\frac{2}{3}$.\label{table:4}}
\end{table*}
\begin{table*}
\begin{tabular}{p{2.5cm}  p{3cm} p{2.5cm} p{4cm} p{2cm} }
\hline
\hline
\multicolumn{5}{c}{Strong Lensing Coefficients }\\
$$Q*2M$$ & $\alpha$& $\bar{a}$  & $\bar{b} $ &$ u_{ph}/R_{sh}$\\
\hline
Schwarzschild&Black Hole&  1.00&-0.40023&2.59808\\
\hline
0&0&  0.999998&-0.400236&2.59806\\
\hline
&0.1& 1.05409&-0.251949&3.04288\\
&0.2&  1.11803&-0.076664&3.63089\\
&0.3&  1.19523&0.134953&4.4361\\
&0.4&  1.1.29099&0.397479&5.59011\\
$5\times 10^{-17}$&0.5&  1.41421&0.735264&7.34838\\
&0.6&  1.58113&1.19286&10.2696\\
&0.7&  1.82573&1.86339&15.8111\\
&0.8&  2.23604&2.98822&29.0465\\
&0.9&  3.16221&5.52718&82.1533\\
\hline
\hline
\end{tabular}
\caption{Estimation  of strong lensing coefficients with the different value of black hole parameters  $\alpha$ and $\gamma*2M=5\times 10^{-17}$   for the case CDM of $SgrA*$ black hole  where $A=6463.81,B=2.537\times10^{-11}$and $\omega=-\frac{2}{3}$.\label{table:5}}
\end{table*}
The behavior of the photon sphere radius $\mathit{r_{ph}}$ in Fig.\ref{fig:8} a and the minimum impact parameter $\mathit{u_{ph}}$ in Fig.\ref{fig:8} b are described as a function of both the parameters $\alpha$ and $\gamma \cdot 2M$ for the case of CDM model. From Fig.\ref{fig:8} a, it is observed that the photon sphere radius $\mathit{r_{ph}}$ increases with the parameters $\alpha$ for the fixed value of $\gamma \cdot 2M$ and also increases with the parameters $\gamma \cdot 2M$ for the fixed value of $\alpha$. From Fig.\ref{fig:8} b and Tables. \ref{table:4} \& \ref{table:5}, it is observed that the minimum impact parameter $\mathit{u_{ph}}$  increases with the parameters $\alpha$ for the fixed value of $\gamma \cdot 2M$ and also increases with the parameters $\gamma \cdot 2M$ for the fixed value of $\alpha$. It also observes that when $A=0$ and $B=0$, $\alpha=0$ and $\gamma \cdot 2M=0$ the photon sphere radius $r_{ph}=1.5$; and the minimum impact parameter $u_{ph}=2.59808$, corresponds to the case of Schwarzschild black hole \cite{Bozza:2002zj}.\\\\
The behavior of deflection limit coefficients $\mathit{\bar{a}}$ in Fig.\ref{fig:9} a and the deflection limit coefficients $\mathit{\bar{b}}$ in Fig.\ref{fig:9} b are described as a function of both the parameters $\alpha$ and  $\gamma \cdot 2M$ for the case of CDM model. From Fig. \ref{fig:9} a and Tables. \ref{table:4} \&  \ref{table:5}, it is observed that deflection limit coefficients $\mathit{\bar{a}}$ increases with the parameters $\alpha$ for the fixed value of $\gamma \cdot 2M$ and also increase with the parameters $\gamma \cdot 2M$ for the fixed value of $\alpha$. From Fig.\ref{fig:9} b and Tables.\ref{table:4} \&  \ref{table:5}, it is observed that deflection limit coefficients $\mathit{\bar{b}}$ increase with the parameters $\alpha$ for the fixed value of $\gamma \cdot 2M$ and also increase with the parameter $\gamma \cdot 2M$ for the fixed value of $\alpha$. It also observe that when $A=0$ and $B=0$,$\alpha=0$ and $\gamma \cdot 2M=0$ the the deflection limit coefficients $\mathit{\bar{a}}=1$; and the deflection limit coefficients $\mathit{\bar{b}}=-0.400023$, corresponds to the case of Schwarzschild black hole \cite{Bozza:2002zj}.\\\\
The behavior of deflection angle $\mathit{\alpha_D}$ in Fig.\ref{fig:10} is described as a function of both the parameters $\alpha$ and $\gamma \cdot 2M$ for the case of the CDM model. It is observed that the deflection angle $\alpha_D$ increases with the parameters $\alpha$ for the fixed value of $\gamma \cdot 2M$ and also increases with the parameters $\gamma \cdot 2M$ for the fixed value of $\alpha$. Moreover the deflection angle $\alpha_D$ for the cases of $M87^{*}$ and $Sgr A^{*}$ exhibit nearly identical behavior. It's worth noting that the photon sphere radius $\mathit{r_{ph}}$, minimum impact parameter $\mathit{u_{ph}}$, deflection limit coefficients $\mathit{\bar{a}}$, deflection limit coefficients $\mathit{\bar{b}}$ and the strong deflection angle $\mathit{\alpha_D}$ for the cases of $M87^{*}$ and $Sgr A^{*}$ exhibit nearly identical behavior for the CDM model.\\\\
The behavior of observable quantities angular image position$ \theta_{\infty}$ in Fig.\ref{fig:11}, image separation $S$ in  Fig.\ref{fig:12}, relative magnification $r_{mag}$ in Fig.\ref{fig:13}  of the relativistic images and the time delays $\Delta T_{2,1}$  between two different relativistic images in  Fig.\ref{fig:14}  are described as a function of both the parameters $\alpha$ and $\gamma \cdot 2M$ for the case of URC model. From Figs.\ref{fig:11} \& \ref{fig:12}, it is observed that observable quantities angular image postion$ \theta_{\infty}$ and image separation $S$  increases with the parameters $\alpha$ for the fixed value of $\gamma \cdot 2M$ and also increase with the parameters $\gamma \cdot 2M$ for the fixed value of $\alpha$ ; but in Fig.\ref{fig:13}, the observable quantity relative magnification $r_{mag}$ decreases with the parameters $\alpha$ for the fixed value of $\gamma \cdot 2M$ and also decreases with the parameters $\gamma \cdot 2M$ for the fixed value of $\alpha$ for both the cases of $M87^{*}$ and $Sgr A^{*}$ supermassive black hole (cf.Table.\ref{table:6}). From Fig.\ref{fig:14} a \& b and Table.\ref{table:6}, it is observed that the time delays $\Delta T_{2,1}$  between two different relativistic images increases with the parameters $\alpha$ for the fixed value of $\gamma \cdot 2M$ and also increases with the parameters $\gamma \cdot 2M$ for the fixed value of $\alpha$ in the context of  CDM halo.
\begin{table*}
\begin{center}
\begin{tabular}{cc|cc|cc|cc|cc}
\hline
\hline
{parameters} && {$ M87^*$}& &{$ Sgr A^*$ }&&{$ M87^*$}&{$ Sgr A^*$}&$ M87^*$  &$Sgr A^*$
\\
$\gamma \cdot 2M$ & $\alpha$ & $ \theta_{\infty} (\mu as)$&$S(\mu as)$&
$\theta_{\infty} (\mu as)$&$S(\mu as)$ &$r_{mag}(magnitude)$&$r_{mag}(magnitude)$& $\Delta T_{2,1}(minutes)$ & $\Delta T_{2,1}(minutes)$ \\
\hline
\hline
{0} & 0&19.9633&0.024984&26.3315&0.0329538&6.82188&6.82188&$17378.8$&11.4973\\
\hline
{0} & 0&19.9631&0.024983&26.3824&0.0330169&6.8219&6.82187&$17378.7$&11.4973\\
\hline
 & 0.1&23.381&0.04746&30.89995&0.0627216&6.47182&6.47179&$20354.2$&13.4658\\
 & 0.2&27.8992&0.094435&36.8705&0.124802&6.1017&6.10166&$24287.4$&16.0679\\
 & 0.3&34.0863&0.198869&45.0471&0.26282&5.70762&5.70758&$29673.5$&19.6312\\
 & 0.4&42.9535&0.4498&56.7657&0.594443&5.28423&5.28419&$37392.8$&24.7381\\
$5\times 10^{-17}$ & 0.5&56.4637&1.11695&74.6203&1.47613&4.82382&4.82378&$49154$&32.519\\
 & 0.6&78.9101&3.15466&104.285&4.16912&4.31456&4.31452&$68694.6$&45.4466\\
 & 0.7&121.489&10.7939&160.556&14.265&3.73653&3.43648&$105762$&69.9692\\
 & 0.8&223.187&51.133&294.957&67.5765&3.05088&3.05081&$194294$&128.54\\
 & 0.9&631.245&497.006&834.239&656.843&2.15732&2.15723&$549525$&363.556\\

\hline
\hline
\end{tabular}
\end{center}
\caption{Estimation of strong lensing observables for CDM case for the supermassive black holes $ M87^*$,$ Sgr A^*$,  with the different value of black hole parameters  $\alpha$; and $\gamma*2M=5\times 10^{-17}$.The observable quantity $r_{mag}$ does not depend on the mass or distance of the black hole from the observer.\label{table:6}}
\end{table*}

\section{Comparison with observation}\label{sec4}

In this paper, we apply the standard method for strong gravitational lensing phenomena
developed by Bozza \cite{Bozza:2002zj} , which can be used for distinguishing between various types of spherically symmetric black holes and exploring the astrophysical implications. The study focuses on supermassive black holes $M87^{*}$ and $SgrA^{*}$  located at the centers of nearby galaxies.  We evaluate the
strong lensing coefficients $\bar{a}$,$\bar{b}$, $u_{Ph}/R_s$ (see,
Tables. \ref{table:1}, \ref{table:2}, \ref{table:4}, \ref{table:5} and \ref{table:7}) and the observable quantities $\theta_{\infty}$, $S$,
$r_{mag}$ (see, Tables. \ref{table:3},\ref{table:6} \&  \ref{table:7}) for the supermassive black hole at the center nearby galaxy. We also evaluate these  observable quantities for of Schwarzschild black hole with halo ($\alpha=0,\gamma*2M$ ) and standard Schwarzschild ($A=0,B=0 ,\alpha=0,\gamma*2M$ ) black hole, black hole space-time with DM halo ($A=805.231$
$B=3.40611*10^{-9} $, $\gamma*2M=1.2\times 10^{-16}$ for URC model; and $A=85.508$
$B=2.38952*10^{-9} $, $\gamma*2M=5\times 10^{-17}$).
by taking the supermassive black hole $M87*$ having mass $M=6.5\times 10^9 M_{\odot}$ and distance $D_{ol}=16.8Mpc$
\cite{EventHorizonTelescope:2019dse} for comparison
(See Table.\ref{VIII}).

\begin{table*}\label{VIII}
\begin{tabular}{ |p{1.6cm}|p{1.5cm}| p{6.5cm}|p{6.4cm}|  }
\hline
& {Schwarz\vfill -schild black hole}&  {  black hole URC model \vfill  $A=805.231$
$B=3.40611*10^{-9} $\vfill($\gamma*2M=1.2\times 10^{-16})$  }\vfill {\begin{tabular}{cccc}
    \hline
    \multicolumn{1}{c}{}\\
    $\alpha=0.2$  & $\alpha=0.4$  &$\alpha=0.6$ &$\alpha=0.8$\\
\end{tabular}} & {black hole CDM model \vfill $A=85.508$
$B=2.38952*10^{-9} $ \vfill($\gamma*2M=5\times 10^{-17})$ }\vfill{\begin{tabular}{cccc}
\hline
    \multicolumn{1}{c}{}\\
     $\alpha=0.2$ & $\alpha=0.4$ &$\alpha=0.6$  &$\alpha=0.8$\\
\end{tabular}}\\
\hline
\textbf {$\theta_{\infty}$\vfill($\mu$arcsecs)} & {14.6901}  &{\begin{tabular}{cccc}
    \hline
    \multicolumn{1}{c}{}\\
    {27.9023}& {42.9599}& {78.9277} & {223.286}\\
\end{tabular}} &{\begin{tabular}{cccc}
    \hline
    \multicolumn{2}{c}{}\\
    {27.9023}& {42.9599}& {78.9277} & {223.287}\\
\end{tabular}}\\
\hline
\textbf {$S$ \vfill($\mu$arcsecs) }&{0.0184}  &{\begin{tabular}{cccc}
    \hline
    \multicolumn{1}{c}{}\\
   { 0.094475}&{0.450029}&{3.154676}&{51.1877}\\
\end{tabular}} &{\begin{tabular}{cccc}
    \hline
    \multicolumn{1}{c}{}\\
    { 0.094475}&{0.450029}&{3.154676}&{51.1877}\\
\end{tabular}}\\
\hline
$r_{mag}$ &{6.82188 } &{\begin{tabular}{cccc}
    \hline
    \multicolumn{1}{c}{}\\
   {6.1017}&  {5.28423}& {4.31456}& {3.05088}\\
\end{tabular}} &{\begin{tabular}{cccc}
    \hline
    \multicolumn{1}{c}{}\\
     {6.1017}&  {5.28423}& {4.31456}& {3.05088}\\
\end{tabular}}\\
\hline

$\Delta T_{2,1}$\vfill{(minutes)} &{17378.8} &{\begin{tabular}{cccc}
    \hline
    \multicolumn{1}{c}{}\\
   { 24290.1}&{37398.4}& {68709.8}& {194380}\\
\end{tabular}} &{\begin{tabular}{cccc}
    \hline
    \multicolumn{1}{c}{}\\
     {24287.4}&{37392.8}& {68694.6}& {194294}\\
\end{tabular}}\\
\hline
{$u_c/R_{sh}$} & {2.59808}  &{\begin{tabular}{cccc}
    \hline
    \multicolumn{1}{c}{}\\
   {3.63129}&{5.59093}& {10.2719}& {29.0592}\\
\end{tabular}}&{\begin{tabular}{cccc}
    \hline
    \multicolumn{1}{c}{}\\
   {3.63089}&{5.5901}& {10.2696}& {29.0463}\\
\end{tabular}} \\
\hline
 {$\bar{a}$ }& {1 } &{\begin{tabular}{cccc}
    \hline
    \multicolumn{1}{c}{}\\
    {1.11807}& {1.29105}&{1.58125}& {2.23637}\\
\end{tabular}}&{\begin{tabular}{cccc}
    \hline
    \multicolumn{1}{c}{}\\
     {1.11803}& {1.29099}&{1.58113}& {2.23604}\\
\end{tabular}}\\
\hline
{$\bar{b}$ }& {-0.40023} &{\begin{tabular}{cccc}
    \hline
    \multicolumn{1}{c}{}\\
    { -0.0765524}& {0.397651}& {1.19318}& {2.98911}\\
\end{tabular}} &{\begin{tabular}{cccc}
    \hline
    \multicolumn{1}{c}{}\\
    { -0.0766659}& {0.397476}& {1.19285}& {2.9882}\\
\end{tabular}}\\
\hline
\end{tabular}
 \caption{Estimation of observables by taking the supermassive black hole $M87^*$  having mass $M=6.5\times 10^9 M_{\odot}$ and distance $D_{OL}=16.8 Mpc$ in the context of  Schwarzschild black hole,  black hole URC model, and black hole CDM model.
 \label{table:7}}
\end{table*}

In our estimation, it is found that
considering the same mass and distance (see Table. \ref{table:7}), the
innermost images $\theta_{\infty}$ in the background of black hole with DM halo(both URC and CDM cases with $\alpha=0.8$) is fifteen times more than
the standard  Schwarzschild black hole. Also, a black hole with a DM halo has a very large image separation $S$ and smaller relative magnification
$r_{mag}$. Their difference value from the standard  Schwarzschild black hole respect as
$\sim 51.17 \mu as$ and $\sim 3.771$ magnitude for $\alpha=0.8$ . It is also observed that for the black hole with a DM halo,
$\theta_{\infty}\in(27.9,223.286)\mu as$ while $S\in(0.09,0.51.2)\mu as$. It
suggests that the outermost images of the black hole with DM halo are closer to the remaining innermost images and which able to be distinguished
from the other black hole images.
Furthermore, if the outermost image can be
resolved, it will distinguish the black hole with a DM halo from Schwarzschild's black hole and characterize the black hole with DM halo by using the current technology.
In Tables \ref{table:3} ,\ref{table:6} \& \ref{table:7}, it is observed that the time delay $\Delta
T_{2,1}$ for the case of  black hole with DM halo
($\gamma*2M=1.2\times 10^{-16}$ ,$\alpha=0.6$) for URC model and ($\gamma*2M=5.7\times 10^{-17}$ ,$\alpha=0.6$)  for CDM model
(e.g.$\sim 68709.8 $ minutes for $M87^{*}$ and $\sim 45.4478 $ minutes for $SgrA^{*}$) is three times more than 
the case of Schwarzschild black hole with halo ($\alpha=0,\gamma*2M$ ) and standard Schwarzschild  ($A=0,B=0 ,\alpha=0,\gamma*2M$ ) black hole. It means
that if one can distinguish the first and second relativistic
images, the time delay between these two images might provide a good chance to detect the black hole with a DM halo from a Schwarzschild black hole. Therefore, the black hole with a DM halo could be quantitatively distinguished from another astrophysical black hole such as a standard Schwarzschild
black holes. It is noted that from the help of
gravitational lensing scenario, it may be easy way to detect an
audible boundary of the sound waves near the black hole horizon under the effetc of DM halos within the scope of string cloud and quintessential field. The astrophysical resulting consequences of the
DM halos exhibiting black hole may shed light on the correlations between sonic fluid and black hole shape. We anticipate that our analytical findings may be useful in the future for observing the comparable black hole research. These findings may facilitate more research into the composition of the near horizon geometry of typical astrophysical black holes.

\section{ Discussions and Conclusions}\label{sec5}
In this paper, we have explored the calculations of strong gravitational lensing and focus on supermassive black holes in the central region of a galaxy, which is surrounded by a DM halos, in the background of Schwarzschild-like spacetimes within the scope of string cloud quintessential field. For the arrangement of two different DM halos namely URC and CDM models, their lapse function have provided in Eq. (\ref{5}) and Eq. (\ref{8}), respectively, we have calculated the values of involved parameters by using the observational data of M87* and Sgr A* viz., $\rho_0$ and $r_0$. We have analyzed the effects of cloud parameter $\alpha$ and DM halo parameter $\gamma \cdot2M$ on the strong gravitational lensing and its observables such as angular image position $ \theta_{\infty}$, image separation $S$ in, relative magnification $r_{mag}$ of the relativistic images and the time delays $\Delta T_{2,1}$  between two different relativistic images compared to the case of standard  Schwarzschild  ($A=0, B=0, \alpha=0, \gamma \cdot 2M=0$), Schwarzschild with halo($\alpha=0 $,$\gamma \cdot 2M=0 $), black hole with URC DM halo model and black hole with CDM halo model. We first calculate the null geodesic equation for the black hole space time by Hamilton -Jacobi
action, and using this, we obtain the photon sphere radius $r_{ph}$.\\\\
It is found that the photon sphere radius $\mathit{r_{ph}}$ increases with the parameters $\alpha$ for the fixed value of $\gamma \cdot2M$ and also increases with the parameters $\gamma \cdot2M$ for the fixed value of $\alpha$  for both the cases of URC and CDM halo models. We obtained strong lensing coefficients   $\mathit{u_{ph}/R_s}$,$\mathit{\bar{a}}$ and $\mathit{\bar{b}}$ numerically as well as graphically for both URC and CDM model. It is observed that strong lensing coefficients   $\mathit{u_{ph}/R_s}$,$\mathit{\bar{a}}$ and $\mathit{\bar{b}}$ are increased with the parameters $\alpha$ for the fixed value of $\gamma \cdot2M$ and also these are increased with the parameters $\gamma \cdot2M$ for the fixed value of $\alpha$  for both the cases of URC and CDM halo models. Using these coefficients, we obtained the deflection angle $\mathit{\alpha_D}$, as a function of both the parameters $\alpha$ and $\gamma \cdot2M$ for both the cases URC  as well as CDM model. It is found that the strong deflection angle $\mathit{\alpha_D}$ increases with the parameters $\alpha$ for the fixed value of $\gamma \cdot2M$ and also increases with the parameters $\gamma \cdot2M$ for the fixed value of $\alpha$  for both the cases of URC and CDM halo models. Thus, both the parameters $\alpha$ and $\gamma \cdot2M$ for the cases of URC and CDM halo greatly intensify the gravitational bending effects.
Furthermore, it is also observed that the strong deflection angle $\mathit{\alpha_D}$  for both the cases of URC and CDM halo models are greater than the case of standard Schwrazschild black hole. The findings indicate that the gravitational lensing effect due to black hole with DM hallo is greatly enhanced compared to the case of standard Schwrazschild black hole. Thus the black hole with DM hallo can be detected more easily and
distinguished from the other ordinary astrophysical black holes such as standard Schwrazschild black holes.
With the help of strong lensing coefficients $\mathit{u_{ph}/R_s}$,$\mathit{\bar{a}}$ and $\mathit{\bar{b}}$, the strong lensing observable quantities such as angular image postion $ \theta_{\infty}$  , image separation $S$, relative magnification $r_{mag}$ of the relativistic images and the time delays $\Delta T_{2,1}$  between two different relativistic images have been obtained for the case of URC  and CDM models.\\\\
We numerically, calculate the strong lensing observables, angular image postion $ \theta_{\infty}$, image separation $S$, relative magnification $r_{mag}$ of the relativistic images and the time delays $\Delta T_{2,1}$ for the cases supermassive black holes $M87^{*}$, $SgrA^{*}$ in the context of black hole spacetime with DM halo. It is seen that the observable quantities angular image postion $ \theta_{\infty}$ and image separation $S$  are increase with the parameters $\alpha$ for the fixed value of $\gamma \cdot2M$ and also increase with the parameters $\gamma \cdot2M$ for the fixed value of $\alpha$  for both the cases of URC and CDM halo models while relative magnification $r_{mag}$ of the relativistic images decreases with the parameters $\alpha$ for the fixed value of $\gamma \cdot2M$ and also decreases with the parameters $\gamma \cdot2M$ for the fixed value of $\alpha$.
In the URC DM halo case where
$\gamma*2M=1.2\times 10^{-16}$and $0\leq \alpha\leq 0.9$, angular position $\theta_{\infty}\in
(19.96,631.9)\mu as$ for $M87^{*}$, $\theta_{\infty}\in
(26.33,834.4)\mu as$ for $SgrA^{*}$ ; the angular separation $S\in (0.024,497.9)\mu
as$ for $M87^{*}$, $S\in (0.032,656.9)\mu as$ for $SgrA^{*}$; and the magnification $r_{mag}\in
(2.1566,6.822) $ magnitude for $M87^{*}$ ,$r_{mag}\in (2.1572,6.822)$ magnitude for $SgrA^{*}$.In the CDM halo case where
$\gamma*2M=5.7\times 10^{-17}$and $0\leq \alpha\leq 0.9$, angular position $\theta_{\infty}\in
(19.96,631.25)\mu as$ for $M87^{*}$, $\theta_{\infty}\in
(26.33,834.24)\mu as$ for $SgrA^{*}$ ; the angular separation $S\in (0.024,497.01)\mu
as$ for $M87^{*}$, $S\in (0.032,656.85)\mu as$ for $SgrA^{*}$; and the magnification $r_{mag}\in
(2.157,6.822) $ magnitude for $M87^{*}$ ,$r_{mag}\in (2.1572,6.822)$ magnitude for $SgrA^{*}$.\\\\
Another important observable quantity, the time delays $\Delta T_{2,1}$  between two different relativistic images increases with the parameters $\alpha$ for the fixed value of $\gamma\cdot2M$ and also increases with the parameters $\gamma \cdot2M$ for the fixed value of $\alpha$  for the cases of URC model as well as CDM model. We have also numerically obtained the time-delays $\Delta T_{2,1}$ between first and second-order relativistic images for the supermassive black holes  $M87^{*}$ and $SgrA^{*}$ in the background of standard   Schwarzschild  ($A=0,B=0,\alpha=0,\gamma \cdot2M=0$ ), Schwarzschild like halo($\alpha=0,\gamma*2M$ ) and black hole with DM halo ($\gamma*2M=1.2\times 10^{-16}$,$0\leq \alpha\leq 0.9$) for URC model and ($\gamma*2M=5.7\times 10^{-17}$,$0\leq \alpha\leq 0.9$).It is observed that the time delay $\Delta
T_{2,1}$ for the case of black hole with DM halo ($\gamma*2M=1.2\times 10^{-16}$,$\alpha=0.6$) for URC model and ($\gamma*2M=5.7\times 10^{-17}$,$\alpha=0.6$)  for CDM model
(e.g.$\sim 68709.8 $ minutes for $M87^{*}$ and $\sim 45.4478 $ minutes for $SgrA^{*}$) is three times more than the case of a Schwarzschild black hole with halo ($\alpha=0,\gamma \cdot2M=0$ ) and standard Schwarzschild    ($A=0, B=0,\alpha=0,\gamma \cdot2M$=0 ) black hole.\\\\
Therefore, the findings in our investigation suggest how the black hole with DM halo under the effect of string cloud quintessential field, is detectable by the astronomical observations. Further, we also try to investigate strong gravitational lensing as extension of the current work by using other DM profiles. It is also our plan to extend our work for the rotating black holes. Finally, we may test the effects of DM halos on various modified theories of gravity.\\

\section*{Acknowledgements}
N.U.M would like to thank  CSIR, Govt. of
India for providing Senior Research Fellowship (No. 08/003(0141))/2020-EMR-I). This research is partly supported by Research Grants FZ-20200929344 and F-FA-2021-510 of the Uzbekistan Ministry for Innovative Development. G. Mustafa is
very thankful to Prof. Gao Xianlong from the Department of Physics, Zhejiang Normal University, for his kind support and help during this research. Further, G. Mustafa acknowledges Grant No. ZC304022919 to support his Postdoctoral Fellowship at Zhejiang Normal University


\bibliographystyle{elsarticle-num}
\bibliography{Main} 


\end{document}